\documentclass[12pt]{article}

\usepackage[totalheight = 23cm, totalwidth = 17cm]{geometry}
\usepackage{amssymb,amsmath,amsfonts,amsbsy,graphicx,bm}

\usepackage{color,cancel,ulem}

\newcommand{\dis}[1]{\begin{equation}\begin{split}#1\end{split}\end{equation}}

\begin{document}

\begin{titlepage}

\begin{center}

{\LARGE \bf 
 Uplift  and  towers of states in   warped throat
}

\vskip 1.0cm

{\large
Min-Seok Seo$^{a}$ 
}

\vskip 0.5cm

{\it
$^{a}$Department of Physics Education, Korea National University of Education,
\\ 
Cheongju 28173, Republic of Korea
}

\vskip 1.2cm

\end{center}

\begin{abstract}

We investigate the connection between the distance conjecture and the uplift potential.
For this purpose, we consider the concrete model, the warped deformed conifold embedded into Type IIB flux compactifications, with  the uplift potential produced by $\overline{\rm D3}$-branes at the tip of the throat.
Whereas the various mass scales associated with towers of states can be found, it turns out that the lightest tower mass scale satisfies the scaling behavior with respect to the uplift potential, which is meaningful provided the number of $\overline{\rm D3}$-branes is nonzero.
This indicates that the effective theory becomes invalid in the vanishing limit of the uplift potential by the descent of an infinite tower of states  from UV, as predicted in the distance conjecture.
Since too large uplift potential is also problematic due to the runaway behavior of the moduli potential as well as the sizeable backreaction of $\overline{\rm D3}$-branes, the uplift potential is bounded from both above and below.
In the simple model like the KKLT or the large volume scenario in which non-perturbative effect is dominatd by the single term, this bound can be rewritten as the bound on the size of the superpotential.

\end{abstract}

\end{titlepage}

\newpage

\section{Introduction}

 The low energy effective  field theory (EFT)  often suffers from the naturalness issues as quantum corrections to a  scalar mass or the  cosmological constant are sensitive to  much higher   energy scale in the absence of some symmetry reason.
 Presumably, they appear problematic because of our ignorance of  quantum gravity, in the context of which the notion of naturalness may change drastically.
 This has recently been one of important topics in the swampland program, which aims to identify  quantum gravity constraints on the low energy EFT in light of   observations in string theory \cite{Vafa:2005ui} (for reviews, see \cite{Brennan:2017rbf, Palti:2019pca, vanBeest:2021lhn, Grana:2021zvf, Agmon:2022thq}).
 
 Many of conjectured criteria distinguishing theories that are consistent with quantum gravity (belong to the `landscape') from that are not (belong to the `swampland') rely on the distance conjecture \cite{Ooguri:2006in}.
 It states that the infinite distance limit of the scalar moduli space corresponds to a particular corner of the landscape, beyond which the EFT breaks down as  an infinite tower of  states descends   from UV.
 A typical example of a tower of states  might be a set of Kaluza-Klein (KK) modes, which would become light if the moduli determining the size of extra dimensions are stabilized at  infinitely large values.

 Concerning the cosmological constant $\Lambda$, the naturalness of an extremely small and positive observed value given by $\sim 10^{-120} m_{\rm Pl}$, where $m_{\rm Pl}=1/\sqrt{8\pi G}$ is the reduced Planck mass, can be studied in the context of the distance conjecture by asking whether the mass scales of towers of states   remain heavy enough to be decoupled from the EFT in the vanishing   limit of $\Lambda$.     
Indeed, it was pointed out that without introducing the negative tension objects,  not only the realization of the de Sitter (dS) and the Minkowski vacuum \cite{Maldacena:2000mw}, but also   the scale separation between  the  KK mass scale  and $|\Lambda|$ in the anti-de Sitter (AdS) vacuum  is challenging   \cite{Gautason:2015tig} in the flux compactifications (see also \cite{Tsimpis:2012tu} for earlier discussion).
Such an observation motivated the  conjecture that for the consistency with quantum gravity, the vanishing limit of $\Lambda$ in   AdS space  corresponds to the infinite distance limit of the moduli space.
Then there exists a tower of states with mass scale $\Delta m$  following the scaling behavior, 
\dis{\Delta m \sim \Big(\frac{|\Lambda|}{m_{\rm Pl}^4}\Big)^\alpha m_{\rm Pl},\label{eq:starp}}
where $\alpha$ is some positive number  \cite{Lust:2019zwm}. 
 Extending this `AdS distance conjecture' (ADC)   to dS space, we can predict the existence of   a tower of light states in the universe with  a small, positive $\Lambda$ as we observe it.
If it is identified with the KK mode,  $\alpha$ is constrained to lie in the range $\frac14 \leq\alpha\leq \frac12$ \cite{Montero:2022prj}, which is obtained  by combining the observational bound on the size of  extra dimensions \cite{Hannestad:2003yd} and the Higuchi bound \cite{Higuchi:1986py}.
We remark here that the breakdown of the EFT in the $\Lambda \to 0$ limit claimed by the ADC does not exclude the  Minkowski vacuum from the landscape. 
 The ADC just tells us the discontinuity between  the Minkowski vacuum with exactly vanishing $\Lambda$ and the (A)dS  vacuum in the $\Lambda \to 0$ limit : they are different branches of the space of vacua in the landscape hence cannot be interpolated by the  EFT consisting of the finite number of fields.

Meanwhile, there are several counterexamples in   string  models allowing the scale separation (see, for example,  \cite{Shiu:2022oti} and references therein).
Moreover, in the language of the low energy effective  supergravity,   the size of  $\Lambda$ is determined by the amount of supersymmetry (SUSY) breaking.
More concretely, if SUSY is unbroken, the universe is in the AdS vacuum with the smallest negative $\Lambda$ (thus the largest $|\Lambda|$) given by $|\Lambda|=3 m_{\rm Pl}^2 m_{3/2}^2$, where $m_{3/2}$ is the gravitino mass.
When SUSY is broken by  F-term, D-term, or the antibrane uplift, the universe can be in the Minkowski or the dS vacuum, as well as the AdS vacuum with smaller $|\Lambda|$.
Then it may be $m_{3/2}$ rather than  $|\Lambda|$ that is connected to a tower of states hence the distance conjecture, as claimed in the `gravitino distance conjecture' \cite{Cribiori:2021gbf, Castellano:2021yye} (see \cite{Palti:2020tsy} for earlier discussion and    \cite{Anchordoqui:2023oqm} for the study on the size of extra dimensions in view of $m_{3/2}$).

 In order to resolve all such ambiguities, we need to investigate  the connection between the mass scale of a tower of states and various ingredients  used to determine $\Lambda$ in  string models, i.e., fluxes, non-perturbative effects, and uplift in more detail.
  For this purpose, we consider the concrete model, the warped deformed conifold supported by background fluxes \cite{Klebanov:2000hb} which is realized in the orientifold  compactifications of Type IIB string theory \cite{Giddings:2001yu}, with the uplift produced by antibranes at the tip of the throat \cite{Kachru:2002gs}.
 We point out in this article that the antibrane uplift which plays the crucial role in realizing the metastable dS vacuum \cite{Kachru:2003aw, Balasubramanian:2005zx}  can be easily connected to the distance conjecture.
 Indeed,  it was already observed in  \cite{Blumenhagen:2022zzw} that when the throat is   strongly warped, both the KK mass scale $m_{\rm KK}$ and the uplift potential $V_{\rm up}$ produced by $\overline{\rm D3}$-branes are redshifted in the same way, satisfying  the scaling behavior given by $m_{\rm KK}\sim (V_{\rm up})^{1/4}$. 
 This gives rise to following questions,  which we try to answer in this article:
 \begin{itemize}
 \item Can we find the similar scaling behavior when the throat is weakly warped? : 
 the scaling behavior in \cite{Blumenhagen:2022zzw} tells us that $m_{\rm KK}$ and $V_{\rm up}^{1/4}$ depend on the stabilized value of the conifold modulus and the volume modulus in the same way. 
 Whereas the value of the conifold modulus does not play the crucial role in the extremely weakly warped throat, the volume dependence still remains, from which we can find out the scaling behavior between $V_{\rm up}$ and the tower mass scale associated with the bulk.
 \item Why $V_{\rm up}$ produced by $\overline{\rm D3}$-branes is directly connected to the tower mass scale? : 
 the tower mass scale like the string or the KK mass scale is determined by the geometry of the internal manifold, such as the size of the throat or the internal volume.
 Meanwhile, the warping of the internal manifold also regulates the size of the four-dimensional spacetime over which $\overline{\rm D3}$-branes are extended, thus the size of $V_{\rm up}$.
 From this, we expect that the tower mass scale and  $V_{\rm up}$ can be connected in a direct way, which will be explored in detail in this article. 
\end{itemize}  
 Moreover, whereas the size of $V_{\rm up}$ is typically identified with the AdS vacuum energy density before uplift, this makes sense only for the tiny cosmological constant as we observe today.
Since they have different origins and can be different in size in the vacuum of   sizeable $|\Lambda|$, we need to distinguish them.
In this sense, in the model building point of view, it is $V_{\rm up}$ rather than $|\Lambda|$ or $m_{3/2}$ which needs to be considered in connection with the distance conjecture.
 Indeed, the exponent $1/4$ in the scaling behavior found in \cite{Blumenhagen:2022zzw} originates from the fact that $\overline{\rm D3}$-branes are   extended over the noncompact four-dimensional  spacetime, which reminds us of the argument in \cite{Montero:2022prj} that the lower bound on $\alpha$ in \eqref{eq:starp} given by $1/4$ is interpreted as the inverse of the number of  noncompact spacetime dimensions. 
 We also find that whereas the lightest tower mass scale   obeys the scaling behavior with respect to $V_{\rm up}$, away from this, there always exists a tower of states satisfying $\alpha=1/4$, even though the mass scale of which may not be   the lightest tower mass scale.

  We emphasize that the scaling behavior   is physically meaningful only if the number of $\overline{\rm D3}$-branes is nonzero. 
  This indicates the discontinuity between the exactly vanishing $V_{\rm up}$ in the absence of $\overline{\rm D3}$-branes and the nonzero but very tiny $V_{\rm up}$   in the following sense.
 Suppose we construct some AdS vacuum by tuning the fluxes and non-perturbative effects but without using the uplift. 
 Here the moduli determining the sizes of the throat and the overall internal volume are stabilized appropriately so that  all possible towers of states are heavy enough not to  affect the low energy EFT. 
  If we try to find the AdS vacuum with the same size of $\Lambda$ using the uplift   in addition, however, the stabilized values of the moduli are strongly restricted not to allow very tiny $V_{\rm up}$ since otherwise there appears a tower of states which becomes extremely light, invalidating the EFT.
  Then we can say that these two AdS vacua with the same size of $\Lambda$ are different branches in the space of vacua. 
  Extension of the argument to  Minkowski or dS space is straightforward : in the moduli space   the Minkowski vacuum stabilized by the fluxes and non-perturbative effects only (see, for example, \cite{Kallosh:2004yh}) is separated from that obtained by the tiny uplift of AdS, in which a tower of states becomes  extremely light as well.  
  Moreover, the uplift is an essential ingredient to realize  the dS vacuum.
 Whereas $V_{\rm up}$ cannot be too large in order not to allow the sizeable backreaction of $\overline{\rm D3}$-branes or the runaway behavior of the moduli potential, our discussion indicates that too tiny $V_{\rm up}$ is also problematic.
 Hence, the size of the AdS cosmological constant before the uplift should not be too small and the Minkowski minimum before the uplift is not allowed when we try to realize the dS vacuum with tiny $\Lambda$ as we observe it.

  The organization of this article is as follows. 
 Section \ref{Sec:WToS} consists of three parts.
 In Section \ref{sec:conifold}, we review the essential features of the warped deformed conifold and discuss the meaning of the strong and weak warping more carefully. 
 In Section \ref{sec:massToS}, we consider the string excitations and the KK modes  as possible towers of states in the string compactifications and present their mass scales in the strongly  and   weakly warped throat, respectively. 
 In  Section \ref{sec:Vupsc}, we investigate whether these mass scales follow the scaling behavior with respect to  $V_{\rm up}$. 
 In order to explore the scaling behavior in more detail, we do not attach  the tuning between the superpotential and $V_{\rm up}$  for the metastable dS vacuum with almost vanishing $\Lambda$.
 Indeed, there is a priori no reason that the  AdS cosmological constant determined by combining various superpotential terms of different origins, namely, fluxes supporting the warped throat, fluxes supporting the bulk, and non-perturbative effects, must be almost the same as $V_{\rm up}$ in size.   
 Meanwhile, as observed in Section \ref{Sec:constraints}, the superpotential and $V_{\rm up}$ are not completely irrelevant, but required to satisfy some inequalities for consistency.
 First, regardless of the sign and the size of $\Lambda$, for the EFT we use to be reliable, it should be protected from the effects of towers of states.
 Therefore, the masses of moduli under consideration must be lighter than the lightest tower mass scale, which imposes the lower bound on $V_{\rm up}$ through the scaling behavior.
 We discuss this constraint by considering the  conifold modulus mass in Section \ref{sec:conmass} and   the gravitino mass as well as the volume modulus mass in Section \ref{sec:m32}, respectively.
 In particular, the condition concerning the gravitino mass imposes the inequality that the superpotential and $V_{\rm up}$ must obey.
 Second, as discussed in Section \ref{sec:run}, in the simple model like the KKLT \cite{Kachru:2003aw} and the large volume scenario \cite{Balasubramanian:2005zx}, $V_{\rm up}$ should not be too large compared to the size of the AdS cosmological constant before uplift, since otherwise the moduli potential shows the runaway behavior and   the  moduli are destabilized.  
 All the discussions above can be rewritten as the constraints on the superpotential :   various terms in the superpotential must be tuned such that when they are summed up, the conditions considered in Section \ref{Sec:constraints} are not violated.
 After emphasizing this, we close our discussion with concluding remarks.
  Appendices are devoted to reviews on Klebanov-Strassler throat, the form of $V_{\rm up}$ describing the brane/flux annihilation, and  the coefficient of the Gukov-Vafa-Witten superpotential, results of which are used throughout this article.
 
 \subsection{Notes on conventions}

Throughout the article, we will focus on  Type IIB superstring theory, on which models like the KKLT \cite{Kachru:2003aw}  or the large volume scenario \cite{Balasubramanian:2005zx} are based.
The string length scale is defined as $\ell_s=2\pi \sqrt{\alpha'}$, the inverse of which $m_s=\ell_s^{-1}$ corresponds to  the string mass scale.
  The bosonic part of the  Type IIB supergravity action is given by
\dis{S_{\rm IIB}=\frac{1}{2 \kappa_{10}^2}\int d^{10}x \sqrt{-G}\Big[{\cal R}-\frac{\partial_M\tau\partial^M\overline{\tau}}{2({\rm Im}\tau)^2}-\frac{g_s|G_3|^2}{2{\rm Im}\tau}-\frac{g_s^2 |F_5|^2}{4}\Big]+\frac{g_s^2}{8 i\kappa_{10}^2}\int\frac{C_4\wedge G_3\wedge \overline{G_3}}{{\rm Im}\tau} \label{eq:TypeIIBaction}} 
 in the Einstein frame, where the string coupling constant $g_s=e^{\Phi_0}\equiv e^{\langle\Phi\rangle} $ is fixed by the dilaton stabilization.
 The metric  is related to that in the string frame by $G_{MN}=e^{-\frac12(\Phi-\Phi_0)}G^S_{MN}$.
Here $\tau=C_0+ie^{-\Phi}$, $G_3=F_3-\tau H_3$, with $F_3=dC_2$ and $H_3=dB_2$, $G_5=dC_4-\frac12 C_2\wedge dB_2+\frac12 B_2\wedge dC_2$, and $|F_p|^2=\frac{1}{p!}F_{M_1\cdots M_p}F^{M_1\cdots M_p}$.
The gravitational coupling in ten-dimensional supergravity is given by $2\kappa_{10}^2=g_s^2(2\pi)^7\alpha^{'4}=g_s^2\ell_s^8/(2\pi)$.

 \section{Connection between tower mass scale and  uplift}
 \label{Sec:WToS}

\subsection{Warped deformed conifold }
\label{sec:conifold}

 To begin with, we consider the ten-dimensional metric  given by
\dis{ds^2=e^{2A(y)}e^{2\Omega(x)}g_{\mu\nu}dx^\mu dx^\nu +e^{-2A(y)}g_{mn}dy^m dy^n,}
where $e^{2\Omega(x)}$ is the Weyl factor that can be chosen freely and $g_{mn}$ is the metric of the Calabi-Yau threefolds in which the deformed conifold (also known as the Klebanov-Strassler throat) is embedded  (see Appendix \ref{App:KSreview} for a review on the throat geometry).
The warp factor $A(y)$ is obtained by solving the equation of motion
\dis{-\widetilde{\nabla}^2 (e^{-4A(y)})=\frac{1}{12{\rm Im \tau}}G_{mnp}\overline{G}^{\widetilde{mnp}}+({\rm localized~source~term}),}
where the tilde in the Laplacian and the upper indices indicates that the metric  $g_{mn}$ rather than $G_{mn}=e^{-2A}g_{mn}$ is used.
Since the equation of motion is invariant under the rescaling $g_{mn}\to \lambda g_{mn}$ and $e^{2A}\to \lambda e^{2A}$ as well as the $y$-independent shift of $e^{-4A}$ \cite{Giddings:2005ff}, one may choose both $\lambda^2$ and the shift to be the same function of $x$, $\sigma(x)$, such that the metric above is rewritten as \cite{Burgess:2006mn, Frey:2008xw}
\dis{ds^2=e^{2A(y)}e^{2\Omega(x)}g_{\mu\nu}dx^\mu dx^\nu+e^{-2A(y)}\sigma(x)^{1/2}g_{mn}dy^m dy^n,\label{eq:metric}}
where the warp factor is given by
\dis{e^{-4A(y)}=1+\frac{e^{-4A_0(y)}}{\sigma(x)},}
which is often denoted by $h(y)$. 
Then $\sigma(x)$ is interpreted as a volume modulus, the stabilization of which fixes the  size of the  overall internal volume.
While $e^{-4A_0}\simeq 0$ in the bulk region,  $e^{-4A_0}$ near the tip of the throat  is given by
\dis{&e^{-4A_0(y)} = 2^{2/3}\frac{(\alpha' g_sM)^2}{\epsilon^{8/3}}I(\eta)
\\
& I(\eta)=\int_\eta^\infty dx \frac{x\coth x-1}{\sinh^2 x}(\sinh(2x)-2x)^{1/3}.\label{eq:KStipwarp}}
  Since the conifold deformation parameter $\epsilon$ has a mass dimension $-3/2$, it is convenient to introduce the dimensionless parameter $|z|=\epsilon^2/\ell_s^3$.

 From $\sqrt{-G}=e^{-2A}e^{4\Omega}\sigma^{3/2}\sqrt{-g_4}\sqrt{g_6}$ and the fact that the four-dimensional part of  the Ricci scalar is given by $e^{-2A}e^{-2\Omega}{\cal R}_4$, one finds that 
\dis{\frac{1}{2 \kappa_{10}^2}\int d^{10}x \sqrt{-G}{\cal R}\supset
\frac{1}{2\kappa_{10}^2}\int d^6y \sqrt{g_6}e^{-4A}\int d^4x e^{2\Omega}\sigma^{3/2}{\color{blue}\sqrt{-g_4}}{\cal R}_4.}
Then it is convenient to choose the Weyl factor to be
\dis{&e^{2\Omega(x)}=\frac{{\cal V}_0 \ell_s^6}{\sigma(x)^{3/2}\int d^6y\sqrt{g_6}e^{-4A}}=\frac{\langle \sigma(x)^{3/2}\rangle}{\sigma(x)^{3/2}},\label{eq:e2Omega}}
where
\dis{ {\cal V}_0 =\frac{\langle \sigma^{3/2}\rangle}{\ell_s^6} \int d^6y\sqrt{g_6}e^{-4A},}
such that $\langle e^{\Omega}\rangle=1$.
We may also rescale the coordinates and $\sigma(x)$ such that $\int d^6y\sqrt{g_6}e^{-4A}=\ell_s^6$ is satisfied hence the internal volume in  units of the string length is simply written as ${\cal V}_0=\langle \sigma^{3/2}\rangle$.
In this case, the gravitational coupling in four dimensions is given by
\dis{ m_{\rm Pl}^2 =\frac{1}{\kappa_4^2}=\frac{{\cal V}_0\ell_s^6}{\kappa_{10}^2},}
which also reads
\dis{m_s=\frac{g_s}{\sqrt{4\pi {\cal V}_0}}m_{\rm Pl}.\label{eq:msmpl}}
This will be used throughout this article to convert the mass scale in terms of $m_s$ into that in terms of $m_{\rm Pl}$.

 Meanwhile, we will observe the behaviors of the potential and the particle spectrum in two limits,  the strongly and weakly warped throat.
 For this purpose, we need to investigate the dominant effects in the   strongly (weakly) warped throat more carefully.
 Through the moduli stabilization,  parameters $\sigma$ and $|z|\equiv \epsilon^2/\ell_s^3$ are fixed at $\langle\sigma\rangle={\cal V}_0^{2/3}$ and $|z|=\Lambda_0^3{\rm exp}[-\frac{2\pi K}{g_sM}]$ \cite{Giddings:2001yu}, respectively.
 Then   we can say the throat is strongly warped if
 \dis{e^{-4A}\simeq \frac{e^{-4A_0}}{\langle\sigma\rangle}\simeq  2^{2/3}I(0)\frac{(g_sM)^2}{(2\pi)^4|z|^{4/3} {\cal V}_0^{2/3}} \gg 1,\label{eq:swnaive}}
  is satisfied (see also \eqref{eq:KStipwarp}).
   When the inequality is reversed, the throat may be said to be weakly warped. 
   One caveat here is that the term containing $e^{-4A_0}/\langle\sigma\rangle \sim (g_sM)^2/[(2\pi)^4|z|^{4/3}{\cal V}_0^{2/3}]$ which we will call the   `warping term' is not always subdominant in the weak warping case defined in this way.
 To see this, we note that the F-term potential for $|z|$ produced by the fluxes is proportional to the inverse of the K\"ahler metric $K^{z\overline{z}}$ (see \eqref{eq:Vforz} and also Appendix \ref{App:conV} for a review) which contains
  \dis{\Big(\log\frac{\Lambda_0^3}{|z|}+\frac{c'(g_s M)^2}{(2\pi)^4{\cal V}_0^{2/3}|z|^{4/3}}\Big)^{-1}.\label{eq:KIzz}} 
  In the $|z| \to 0$ limit, the condition \eqref{eq:swnaive} is satisfied, and   the term in the parenthesis is evidently dominated by the second  warping term.
 Meanwhile, in the opposite limit $|z|/\Lambda_0^3 \to 1$, the  warping term again dominates over the first logarithmic term $\log({\Lambda_0^3}/{|z|})$ which can be identified with the throat length $\eta_{\rm UV}$ (see discussion below \eqref{eq:e4Asw})  as $\eta_{\rm UV}$ becomes close to $0$.
 Indeed, the combination $|z|^{2/3}[\log(\Lambda_0^3/|z|)]^{1/2}$ is maximized at $|z|/\Lambda_0^3=e^{-3/4}\simeq 0.47$,  having the value $[3^{1/2}/(2 e^{1/2})]\Lambda_0^2\simeq 0.53\Lambda_0^2$.
 Then the logarithmic term dominates over the warping term when $|z|$  is in the range around $|z|=e^{-3/4}\Lambda_0^3$, where the lower(upper) bound becomes closer to zero (one) when ${\cal V}_0$ gets larger.

  If we restrict our attention to  the case in which $|z|$ is so small that $|z|^{2/3}{\cal V}_0^{1/3} \ll (g_sM)/(2\pi)^2$ is satisfied (the strong warping in the sense of \eqref{eq:swnaive}),  we may define the strongly warped throat in a more restrictive way by  imposing  
   \dis{\frac{(2\pi)^2|z|^{2/3}{\cal V}_0^{1/3}}{g_s M}\Big(\log\frac{\Lambda_0^3}{|z|} \Big)^{1/2}=\frac{(2\pi)^2|z|^{2/3}{\cal V}_0^{1/3}}{g_s M}\eta_{\rm UV}^{1/2} \ll 1,\label{eq:sscond}}
   i.e., the dominance of the warping term over the logarithmic term in $K^{z\overline{z}}$,    as considered in \cite{Bento:2021nbb}.    
  That is, the upper bound on the combination $|z|^{2/3}{\cal V}_0^{1/3}$ in the strongly warped throat is more restricted by the factor $1/\eta_{\rm UV}^{1/2}$, which is  smaller than $1$ for $|z|/\Lambda_0^3\ll 1$.
  Then the throat in which
  \dis{\frac{1}{\eta_{\rm UV}^{1/2}}<\frac{(2\pi)^2|z|^{2/3}{\cal V}_0^{1/3} }{g_sM}<1\label{eq:wwcond}}  
is satisfied also belongs to the weakly warped throat.
In the same way, we can also divide the case of the `extremely weakly warped throat', in which  the warping term $[(2\pi)^2|z|^{2/3}{\cal V}_0^{1/3}]/(g_sM)$ is similar to or even larger than $1$ thus $e^{-4A}$ takes the value around $1$, into two classes : a)  the warping term is subdominant compared to  $\eta_{\rm UV}$   in $K^{z\overline{z}}$ as $|z|/\Lambda_0^3$ is still smaller than $1$ but  ${\cal V}_0$ is large (thus satisfying $(2\pi)^2|z|^{2/3}{\cal V}_0^{1/3}\gtrsim (g_sM)/\eta_{\rm UV}^{1/2}$), and b) $|z|/\Lambda_0^3$ is too close to $1$ so the warping term dominates over the logarithm term in $K^{z\overline{z}}$.
In this article, we mainly focus on the strongly (weakly) warped throat in the sense of \eqref{eq:sscond} (\eqref{eq:wwcond}) and discuss the `extremely weakly warped throat'  separately.

\subsection{Mass scales of towers of states}
\label{sec:massToS}

 In this section, we explore the possible towers of states in the compactification of Type IIB string theory and their mass scales in the presence of the warped deformed conifold.
 First of all, the string excitations produce a tower of states with the mass scale $m_s$ given by  \eqref{eq:msmpl}.
Moreover, compactifying the ten-dimensional theory  on six-dimensional   manifold  naturally introduces a tower of states consisting of the KK modes.
From the Laplacian associated with the metric \eqref{eq:metric},
\dis{\nabla^2&=\frac{1}{\sqrt{-G}}\partial_M (\sqrt{-G}G^{MN}\partial_N)
\\
&=\frac{1}{e^{2A(y)}e^{2\Omega(x)}}\Big[\frac{1}{e^{2\Omega(x)}\sigma(x)^{3/2}}\partial_\mu(e^{2\Omega(x)}\sigma(x)^{3/2}g^{\mu\nu}\partial_\nu)+\frac{e^{2\Omega(x)}}{e^{-4A(y)}\sigma(x)^{1/2}} \frac{1}{\sqrt{g_6}}\partial_m(\sqrt{g_6}g^{mn}\partial_n)\Big],\label{eq:laplacian}}
and the facts that $\langle e^{\Omega}\rangle=1$ and ${\cal V}_0=\langle \sigma^{3/2}\rangle$, one finds that the KK mass scale is given by
\dis{m_{\rm KK}^2=\frac{\langle e^{2\Omega}\rangle}{\langle e^{-4A}\rangle \langle\sigma^{1/2}\rangle}\frac{1}{R^2}=\frac{1}{\langle e^{-4A}\rangle {\cal V}_0^{1/3}}\frac{1}{R^2},\label{eq:KK}}
where   $R$ is the typical length scale of the internal manifold.

For the KK modes  in the bulk, $\langle e^{-4A}\rangle=1$ and $2\pi R=\ell_s = m_s^{-1}$ are taken, then their mass scale is estimated as
\dis{m_{\rm KK}=\frac{2\pi m_s}{{\cal V}_0^{1/6}} =\sqrt{\pi}\frac{g_s}{{\cal V}_0^{2/3}}m_{\rm Pl}. \label{eq:bulkKK}}
On the other hand, the mass scale of the KK modes localized near the tip of the throat is redshifted by the warp factor.
Near the tip, the deformed throat is equivalent to $S^3 \times S^2\times \mathbb{R}$ (see  \eqref{eq:KStipmetric}) and the typical length scale  of the $S^3$ part (A-cycle) is given by $R^2=\epsilon^{4/3}(2/3)^{1/3}=\ell_s^2|z|^{2/3}(2/3)^{1/3}$.
For the $S^2\times \mathbb{R}$ part (B-cycle), the throat length  scale $\eta_{\rm UV}$ is multiplied to $R$ in addition, giving  $R^2/\ell_s^2=\eta_{\rm UV}^2|z|^{2/3}(2/3)^{1/3}$ \cite{Blumenhagen:2019qcg}.
 Since we are primarily interested in the case of  $\langle e^{-4A}\rangle \simeq \langle e^{-4A_0}/\sigma \rangle\simeq 2^{2/3}I(0) (g_sM)^2/[(2\pi)^4 |z|^{4/3}{\cal V}_0^{2/3}]$, we focus on the lowest KK mass scale in the presence of the warped throat  given by
 \footnote{More precisely, $m_{\rm KK}^{\rm w}/m_{\rm KK}\sim [|z|^{1/3}{\cal V}_0^{1/3}]/[\eta_{\rm UV}(g_sM)]$, which is smaller than ${\cal V}_0^{1/6}/[(2\pi)(g_sM)^{1/2}\eta_{\rm UV}]$.
 On the other hand, the condition $g_s M>(2\pi)^2$ can be imposed in addition from the requirement that the squared length scale of the deformed conifold $\alpha' g_sM$ (see \eqref{eq:sw10Dmetric}) is larger than $\ell_s^2$ for the metric as a supergravity solution to be a valid description \cite{Blumenhagen:2019qcg}.
 Then $m_{\rm KK}^{\rm w}/m_{\rm KK}<{\cal V}_0^{1/6}/[(2\pi)^2\eta_{\rm UV}]$.
 Since even large volume ${\cal V}_0\sim 10^3$ gives ${\cal V}_0^{1/6} \sim \pi$, the upper bound on the ratio is typically smaller than $1$.
\label{foot1} }
\dis{m^{\rm w}_{\rm KK}=\frac{2^{1/2}3^{1/6} \pi^{3/2}}{I(0)^{1/2}} \frac{|z|^{1/3}}{M \eta_{\rm UV} {\cal V}_0^{1/3}}m_{\rm Pl} \sim \frac{|z|^{1/3}}{M \eta_{\rm UV}  {\cal V}_0^{1/3}}m_{\rm Pl}.\label{eq:KKsw}}

We now consider the  extremely weakly warped throat, in which $[(2\pi)^2|z|^{2/3}{\cal V}_0^{1/3}]/(g_sM)$ is similar to or larger than $1$ such that $e^{-4A}$ is close to $1$.
Even in this case, $R^2/\ell_s^2$ is still given by $\eta_{\rm UV}^2|z|^{2/3}(2/3)^{1/3}$,  then the mass scale of the KK modes localized in the throat is given by
\dis{m_{\rm KK}^{\rm ew} \sim \frac{g_s}{\eta_{\rm UV}|z|^{1/3} {\cal V}_0^{2/3}}m_{\rm Pl}=\frac{g_s |z|}{\eta_{\rm UV}|z|^{4/3} {\cal V}_0^{2/3}}m_{\rm Pl}.\label{eq:mKKew}}
Imposing the extremely weak warping condition, the upper bound on $m_{\rm KK}^{\rm ew}/m_{\rm Pl}$ is given by $(2\pi)^4 g_s|z|/[\eta_{\rm UV}(g_sM)^2]$, which is again smaller than $g_s|z|/\eta_{\rm UV}$ since $g_sM>(2\pi)^2$. 
When $\eta_{\rm UV}$ is larger than the warping term, we also have another  bound $m_{\rm KK}^{\rm ew}/m_{\rm Pl}<(2\pi)^4 g_s|z|/ (g_sM)^2 <g_s|z|$, which can be stronger than the previous bound for $\eta_{\rm UV}$ smaller than $1$.  
Thus, $m_{\rm KK}^{\rm ew}$ can be sub-Planckian.
On the other hand, as $|z|/\Lambda_0^3$ gets close to $1$, $\eta_{\rm UV}\to 0$,  then $m_{\rm KK}^{\rm ew}$ becomes extremely heavy, invalidating the EFT.
Indeed, since  $m_{\rm KK}^{\rm ew}/m_{\rm KK}\sim 1/[|z|^{1/3}\eta_{\rm UV}]$, $m_{\rm KK}^{\rm ew}$ is typically heavier than   the bulk KK mass scale $m_{\rm KK}$ : for $|z|/\Lambda_0^3 < 1$, $\eta_{\rm UV}=\log(\Lambda_0^3/|z|)$ varies mildly with respect to $|z|$ compared to $|z|^{1/3}$  and for $|z|/\Lambda_0^3 \simeq  1$, $\eta_{\rm UV}=0$. 
Then  the bulk KK scale $m_{\rm KK}$ is typically the lowest tower mass scale.
 

It is worth   observing how the structure of the metric is reflected in that of the KK mass scale given by \eqref{eq:KK}.
 Expressing the ten-dimensional metric in a more comprehensive way,
 \dis{ds^2=e^{2\Omega_4(x, y)}g_{\mu\nu}dx^\mu dx^\nu+e^{2\Omega_6(x, y)}g_{mn} dy^m dy^n,}
 where
 \dis{e^{2\Omega_4(x, y)}=e^{2A(y)}e^{2\Omega(x)},\quad 
 e^{2\Omega_6(x, y)}=e^{-2A(y)}\sigma(x)^{1/2},} 
one immediately infers from \eqref{eq:laplacian} that the KK mass scale is written as 
\dis{m_{\rm KK}=\langle e^{\Omega_4}\rangle\frac{1}{\langle e^{\Omega_6}\rangle R}=\langle e^{A}\rangle\frac{1}{\langle e^{\Omega_6}\rangle R},}
where $R\sim \eta_{\rm UV} \epsilon^{2/3} =\eta_{\rm UV} |z|^{1/3} \ell_s$.  
This evidently shows that $\langle e^{\Omega_4}\rangle = \langle e^{A}\rangle$ corresponds to the redshift factor.
 When the throat is warped satisfying $|z|^{2/3}{\cal V}_0^{1/3} \ll (g_sM)/(2\pi)^2$, $e^{\Omega_6}$ becomes independent of ${\cal V}_0$ as $e^{\Omega_6} \simeq (e^{-A_0}/\sigma^{1/4})\times \sigma^{1/4}\sim (g_s M)^{1/2} |z|^{-1/3}$.
Moreover, $m_{\rm KK}$ is inversely proportional to $\langle e^{\Omega_6}\rangle R$, in which the factor $|z|^{-1/3}$ in $e^{\Omega_6}$ is cancelled by the factor $|z|^{1/3}$ in $R$ (see discussion above \eqref{eq:sw10Dmetric}). 
Then the KK     mass scale is estimated as $m_{\rm KK}^{\rm w} \propto   \langle e^{A}\rangle  m_s \sim |z|^{1/3}{\cal V}_0^{1/6}  m_s$, which   coincide with \eqref{eq:KKsw}.
In contrast, when the throat is extremely weakly warped, $e^A$  is no longer proportional to ${\cal V}_0^{1/6}|z|^{1/3}$ so the factors $\langle \sigma^{1/2}\rangle={\cal V}_0^{1/6}$ in $e^{\Omega_6}$ and $|z|^{1/3}$ in $R$ are not  cancelled.
Since $\langle e^{\Omega_4}\rangle=1$, one finds that $m_{\rm KK}^{\rm ew}\propto |z|^{-1/3}{\cal V}_0^{-1/6}  m_s$.



 \subsection{Uplift potential and a tower of states} 
 \label{sec:Vupsc}
 
 We now put $\overline{\rm D3}$-branes at the tip of the throat.  
 This breaks SUSY, and can be used to uplift the potential for the volume modulus $\sigma(x)$ to a metastable dS vacuum.
 The uplift potential is given by the sum of the DBI action and the Chern-Simons term.
 Since these two are the same in magnitude, we obtain $V_{\rm up}= 2 p (T_3/g_s)\int d^4 x\sqrt{-\gamma}$, where $p$ is the number of $\overline{\rm D3}$-branes, $T_3=2 \pi \ell_s^{-4}$ is the D$3$-brane tension, and $\gamma$ is the induced metric on $\overline{\rm D3}$-branes.
 If $\overline{\rm D3}$-branes are extended over the noncompact four-dimensional spacetime, the induced metric is given by $ds_{\overline{D3}}^2=e^{2\Omega_4(x,y)}g_{\mu\nu}dx^\mu dx^\nu=e^{2A(y)}e^{2\Omega(x)}g_{\mu\nu}dx^\mu dx^\nu$,  which gives
 \dis{V_{\rm up}= 2 p \frac{T_3}{g_s}e^{4\Omega_4(x, y)}=4\pi p\frac{m_s^4}{g_s}e^{4A(y)}e^{4\Omega(x)},\label{eq:uplift}}
 where   $e^{4A}$ corresponds to the redshift factor.
We note that the factor $4$ in the exponent of $e^{4\Omega_4(x, y)}$ comes from the four noncompact dimensions over which $\overline{\rm D3}$-branes are extended.
 As sketched in Appendix \ref{App:NS5}, the same result is obtained in the context of the brane/flux annihilation, which is described by the polarization of  NS5-brane wrapping the $S^3$ part of the throat \cite{Kachru:2002gs, Brown:2009yb}.

When the throat is warped satisfying $|z|^{2/3}{\cal V}_0^{1/3} \ll (g_sM)/(2\pi)^2$,  $V_{\rm up}$ is estimated to be
\dis{V_{\rm up}^{\rm  w}= \frac{2^{4/3}\pi^3}{I(0)}\frac{g_s p}{M^2}\frac{|z|^{4/3}}{\sigma(x)^2} m_{\rm Pl}^4,\label{eq:upsw}}
or $\langle V_{\rm up}^{\rm w} \rangle \sim (g_s p/M^2)(|z|^{4/3}/{\cal V}_0^{4/3}) m_{\rm Pl}^4$.
As pointed out in \cite{Blumenhagen:2022zzw}, $\langle V_{\rm up}^{\rm  w}\rangle $ depends on ${\cal V}_0$ and $|z|$ in the same way as $(m_{\rm KK}^{\rm  w})^4$.
More concretely,  comparing \eqref{eq:upsw} with \eqref{eq:KKsw} we obtain the scaling behavior
\dis{m_{\rm KK}^{\rm  w} \sim \frac{1}{g_s^{1/4}\eta_{\rm UV}M^{1/2} p^{1/4}} \langle V_{\rm up}^{\rm  w}\rangle^{1/4}.\label{eq:mKKsw}}
 We note that in addition to the power-law dependence on $|z|$ and ${\cal V}_0$ which is relevant to the scaling behavior, $m_{\rm KK}^{\rm  w}$ also contains  the logarithmic term   $\eta_{\rm UV}= \log(\Lambda_0^3/|z|)$.
Indeed, whereas $V_{\rm up}$ is generated by  $\overline{\rm D3}$-branes localized at the tip of the throat and redshifted by $e^{4A}$, the KK mode is   determined by the overall size of the throat, $\eta_{\rm UV}$, with the same redshift effect.

Such a simple scaling behavior also appears  in the extremely weakly warped throat, but the associated tower mass scale is not the throat KK mass scale.
To see this, we compare  $\langle V_{\rm up}\rangle \sim (p/g_s)\langle e^{4A}\rangle m_s^4$ obtained from \eqref{eq:uplift} ($\langle e^{\Omega(x)}\rangle=1$ is used) with the throat KK mass scale given by \eqref{eq:KK},    $m_{\rm KK}^4\sim \langle e^{4A}\rangle R^{-4} \sim \langle e^{4A}\rangle(\langle e^{\Omega_6}\rangle|z|^{1/3}/\eta_{\rm UV})^4 m_s^4$.
 We have seen that as the warping gets stronger, the combination $\langle e^{\Omega_6}\rangle|z|^{1/3}=\langle e^{-A}\rangle {\cal V}_0^{1/6}|z|^{1/3}$ becomes independent of both ${\cal V}_0$ and $|z|$ hence away from $\eta_{\rm UV}$, $m_{\rm KK}^4$ depends on ${\cal V}_0$  and $|z|$ through the combination $\langle e^{4A}\rangle m_s^4$, just like $\langle V_{\rm up}\rangle$. 
  In contrast, when we estimate the throat KK mass scale for the extremely weakly warped throat,  $e^{-A}\simeq 1$ does not cancel ${\cal V}_0^{1/6}$  and $|z|^{1/3}$ any longer.
Indeed, the uplift potential in the extremely weakly warped throat is written as
 \dis{V_{\rm up}^{\rm ew}=
 \frac{g_s^3}{4\pi}\frac{p}{\sigma(x)^3}m_{\rm Pl}^4\Big[1+\frac{2^{2/3}(g_sM)^2I(0)}{(2\pi)^4|z|^{4/3} \sigma(x)}\Big]^{-1}
 \simeq \frac{g_s^3}{4\pi}\frac{p}{\sigma(x)^3}m_{\rm Pl}^4,\label{eq:ewuplift}}
 such that $\langle V_{\rm up}^{\rm ew}\rangle/m_{\rm Pl}^4 =[g_s^3/(4\pi)]p/{\cal V}_0^2$.
 While this expression presumes that  $\overline{\rm D3}$-branes are localized at the tip of the throat, this in fact is not well guaranteed in the extremely weakly warped throat.
 Indeed, the position of $\overline{\rm D3}$-branes in the throat can be found from the  value of $\eta$  at which   $V_{\rm up}^{\rm ew}$ with the $\eta$ dependence restored,
 \dis{\frac{g_s^3}{4\pi}\frac{p}{\sigma(x)^3}m_{\rm Pl}^4\Big[1+\frac{2^{2/3}(g_sM)^2I(\eta)}{(2\pi)^4|z|^{4/3} \sigma(x)}\Big]^{-1},}
 is stabilized.
 Since this is a monotonically increasing function of $\eta$,  $\overline{\rm D3}$-branes are stabilized at $\eta=0$ classically.
 On the other hand, as the throat is extremely weakly warped such that the warping term is much smaller than one for any value of $\eta$, the increasing rate is also suppressed, so in terms of $\eta$, the position of $\overline{\rm D3}$-branes in a throat, $V_{\rm up}^{\rm ew}$ corresponds to the very shallow potential.
 Then quantum mechanically, the probability that $\overline{\rm D3}$-branes are located in other region of the throat, or even outside the throat is not negligible.
 Nevertheless, as we can learn from the basic quantum mechanics example, the bound state must exist even when the depth of the potential is very tiny, in which  the probability to find the $\overline{\rm D3}$-branes inside the potential is still larger than that to find $\overline{\rm D3}$-branes outside the throat.
 Moreover, we also expect that the probability is maximized at $\eta=0$, the tip of the throat, since this is the point at which   $V_{\rm up}^{\rm ew}$ is stabilized classically.
 This may  become invalid if $|z|/\Lambda_0^3$ becomes close to one, or equivalently, $\eta_{\rm UV} \simeq 0$, in which the throat region terminates even before the potential increases in a meaningful size so the dominance of the probability to find   $\overline{\rm D3}$-branes at the tip of the throat is not so strong. 
 Therefore,  in the following discussion, we focus on the case in which $\eta_{\rm UV}$ is still sizeable so the localization of $\overline{\rm D3}$-branes  at the tip of the throat is relatively reliable.

 From \eqref{eq:ewuplift}, one finds that  $V_{\rm up}^{\rm ew}$ depends on $\sigma(x)$ through $e^{4\Omega(x)}={\cal V}_0^2/\sigma(x)^3$ only, so unlike  $V_{\rm up}^{\rm  w}$ which is proportional to $\sigma^{-2}$, $V_{\rm up}^{\rm ew} \propto \sigma^{-3}$.
 We note that since $\langle e^\Omega\rangle=1$, after the stabilization of $\sigma(x)$,  $\langle V_{\rm up}^{\rm ew}\rangle$ can be written as $4\pi p (m_s^4/g_s)$, which evidently shows that $\langle V_{\rm up}^{\rm ew}\rangle$ is independent of $|z|$.
 This explains why $\langle V_{\rm up}^{\rm ew}\rangle$ is not simply related to $m_{\rm KK}^{\rm ew}$ or $m_{\rm W}^{\rm ew}$ through the scaling behavior.
 Instead, we can find two possible towers of states which satisfy the scaling behavior with respect to  $\langle V_{\rm up}^{\rm ew}\rangle $.
 The first one is the string excitations : the sting mass scale $m_s$ satisfies the scaling behavior given by
 \dis{m_s \sim \Big(\frac{g_s}{4\pi p}\Big)^{1/4} \langle V_{\rm up}^{\rm ew}\rangle^{1/4}.} 
 This reflects the fact that  when we fix $m_{\rm Pl}$,  $m_s$ becomes light  in the  ${\cal V}_0 \to \infty$ limit.
 \footnote{We note that $m_s$ is the fundamental mass scale from which $m_{\rm Pl}$ is induced through compactification.
Hence, the precise statement is that $m_s$ admitting $m_{\rm Pl}\sim 10^{18}$GeV and the ${\cal V}\to\infty$ limit is very light.
 }
Indeed, as $m_s \to 0$, the D-brane tension also decreases in size, which makes   $\langle V_{\rm up}^{\rm ew}\rangle$ given by the energy stored in  $\overline{\rm D3}$-branes smaller.
Another one is  the bulk KK mass scale given by \eqref{eq:bulkKK}  satisfying
\dis{m_{\rm KK}\sim \frac{1}{p^{1/3}}\Big\langle\frac{V_{\rm up}^{\rm ew}}{m_{\rm Pl}^4}\Big\rangle^{1/3}m_{\rm Pl}.}
This is not strange because $m_{\rm KK}$ becomes light as ${\cal V}_0$ increases.
In any case, towers of states satisfying the scaling behavior with respect to $\langle V_{\rm up}^{\rm ew}\rangle$ are relevant to the overall internal volume ${\cal V}_0$, not  the throat geometry.
Moreover, as we discussed in Section \ref{sec:massToS}, in the extremely weakly warped throat,  $m_{\rm KK}$ is typically lighter than $m_{\rm KK}^{\rm ew}$, so we can say that in both strongly and (extremely) weakly warped throat, the lightest tower mass scale obeys the scaling behavior with respect to $V_{\rm up}$.

 It is remarkable that we can always find  a tower of states satisfying  $\Delta m \sim V_{\rm up}^{1/4}$, or more precisely, $\Delta m \sim V_{\rm up}^{1/d}$, where $d$ is the number of noncompact spacetime dimensions.
 Here $\Delta m$ corresponds to the KK mass scale for the warped throat satisfying $|z|^{2/3}{\cal V}_0^{1/3} \ll (g_sM)/(2\pi)^2$ and the string mass scale for the extremely weakly warped throat.
We also emphasize that the scaling behavior makes sense only if  the number of $\overline{\rm D3}$-branes is nonzero, i.e., $p\ne 0$.
 Indeed, the existence of a tower of states is a priori irrelevant to  the presence of the uplift.
 In the absence of the uplift, i.e., when $p=0$, the spacetime geometry is given by the AdS vacuum, which is a well  defined four-dimensional supergravity solution  so far as  $\Delta m$ is well separated from the gravitino mass scale or the masses of the moduli under consideration.
 In the presence of the uplift  potential, the scaling behavior becomes meaningful and indicates that the   vacuum constructed by taking  the $\langle V_{\rm up}\rangle \to 0$ limit corresponds to the infinite distance limit of the moduli space.
 That is,   the values of the stabilized moduli consistent with the $\langle V_{\rm up}\rangle \to 0$ limit allows the tiny tower mass scale, indicating the descent of a tower of states from UV as claimed in the distance conjecture.

\section{Constraints on superpotential and uplift}
\label{Sec:constraints}

 Whereas the potential produced by the fluxes and non-perturbative effects which we will  denote by  $V_{\rm AdS}$ stabilizes the volume modulus $\sigma(x)$ in the AdS minimum (or possibly, the (meta)stable Minkowski vacuum) with $|\Lambda|$ given by $|\langle V_{\rm AdS}\rangle|\equiv |V_{\rm AdS}|$, the dS or the Minkowski vacuum as well as the AdS vacuum with smaller $|\Lambda|$  can be realized by adding the uplift  potential $V_{\rm up}$ to   $V_{\rm AdS}$.
 As we have seen in the previous section, the mass scale of a tower of states obeys the scaling behavior with respect to $V_{\rm up}$, implying that  when  $V_{\rm up}$ becomes too small, a tower of states descends from UV, invalidating the EFT.
 In other words, the AdS vacuum determined by $V_{\rm AdS}$ cannot be perturbed by the tiny uplift effect.
 On the other hand,  $V_{\rm up}$ is required not to be too larger  than the sum of $|V_{\rm AdS}|$ and $V_h$, the height of the potential $V_{\rm AdS}$ at the local maximum,  since otherwise the combined potential $V_{\rm AdS}+V_{\rm up}$ does not have local minima but shows the runaway behavior.
Indeed, when $V_{\rm up}$ becomes too large, the backreaction of antibranes on the background geometry is no longer negligible.
  The sum $|V_{\rm AdS}|+V_h$ is more or less comparable to $|V_{\rm AdS}|$ in the simple models like the KKLT or the large volume scenario, where the non-perturbative effect is dominated by the single term (in fact, in the KKLT scenario, $V_{\rm AdS}$ does not have a local maximum).
 In this case, while $V_{\rm up}$ almost comparable to $|V_{\rm AdS}|$ can be used to realize the Minkowski or the (A)dS vacuum with tiny $\Lambda$, it should not be much larger than $|V_{\rm AdS}|$, say, ${\cal O}(10)\times |V_{\rm AdS}|$.   
 In the presence of more than two comparable non-perturbative effects, tuning between them may allow the sum $|V_{\rm AdS}|+V_h$ much larger than $|V_{\rm AdS}|$.
 In this article, we restrict our attention to the simple case where the condition $V_{\rm up}\lesssim {\cal O}(10)\times |V_{\rm AdS}|$ is imposed.
 We also note that whereas the models we are considering aim to realize the metastable dS vacuum, we do not attach this goal but allow the AdS and the Minkowski vacuum, as well as the dS vacuum with the sizeable $\Lambda$.
 From this, we investigate the range of the superpotential consistent with the bounds on $V_{\rm up}$   for the valid EFT description.


 Meanwhile, the size of $|V_{\rm AdS}|$   cannot be  larger than the supersymmetric vacuum energy given by $\Lambda_{\rm SUSY}=3 m_{\rm Pl}^2 m_{3/2}^2$, where $m_{3/2}=e^{K/(2m_{\rm Pl}^2)}|W|/m_{\rm Pl}^2 \sim |W|/(m_{\rm Pl}^2{\cal V}_0)$ is the gravitino mass.
 Since $V_{\rm AdS}$ is the F-term potential produced by the fluxes and non-perturbative effects, the size of $|V_{\rm AdS}|$ is determined by the amount of SUSY breaking parametrized by  the F-term, $F^a=e^{K/(2 m_{\rm Pl}^2)}K^{a\overline{b}}D_{\overline b}\overline{W}$.
  In the minimal model of the KKLT scenario \cite{Kachru:2003aw},  the F-term vanishes, so $V_{\rm AdS}$ stabilizes $\sigma(x)$  in the supersymmetric AdS minimum satisfying $\langle V_{\rm AdS} \rangle=-\Lambda_{\rm SUSY}$.
 In this case, SUSY is broken  by $V_{\rm up}$ only.
   In the large volume scenario \cite{Balasubramanian:2005zx}, on the other hand, whereas  the largest mass scale of the volume modulus is around $m_{3/2}$ \cite{Conlon:2005ki}  the F-term has nonzero vacuum expectation value, so  $| V_{\rm AdS}|\sim (|W|^2/m_{\rm Pl}^2)(\log {\cal V}_0/{\cal V}_0^3)$ is suppressed compared to $\Lambda_{\rm SUSY}$ by  $\log {\cal V}_0/{\cal V}_0$. 
 Then we can impose the minimal condition on  $V_{\rm up}$ and the superpotential in the simple model like the KKLT or  the large volume scenario given by
  \dis{&\Delta m > m_{3/2},\quad\quad
  {\rm and}
  \\
  &V_{\rm up}\lesssim {\cal O}(10)\times |V_{\rm AdS}|\leq {\cal O}(10)\times \Lambda_{\rm SUSY}={\cal O}(10)\times 3 m_{\rm Pl}^2 m_{3/2}^2,\label{eq:bounds}}
  which may be more restricted depending on the model.
  For instance, the lower bound on $\Delta m$ is expected to be the heaviest mass of the moduli under consideration. 
  We note that if the non-perturbative effects are tuned such that $|V_{\rm AdS}|+V_h$ becomes much larger than $|V_{\rm AdS}|$, it introduces the mass scale $m_h$ defined by   $|V_{\rm AdS}|+V_h= m_{\rm Pl}^2 m_h^2$.
  Then one may  expect that the volume modulus mass $\sigma(x)$  is not much enhanced compared to $m_h$, from which the constraint in the form of  \eqref{eq:bounds} with $m_{3/2}$ replaced by $m_h$ can be imposed.
  This is quite model dependent so we do not explore in more detail.
  
  As we have seen in Section \ref{sec:Vupsc}, we can always find a tower of states obeying the scaling behavior with respect to $V_{\rm up}$.
  When the throat is warped satisfying $|z|^{2/3}{\cal V}_0^{1/3} \ll (g_sM)/(2\pi)^2$, the  throat KK mass scale   $m_{\rm KK}^{\rm w}$ is   the lowest tower mass scale and at the same time, scales like $m_{\rm KK}^{\rm w}\sim \langle V_{\rm up}^{\rm w}\rangle^{1/4}$.
  Then the first condition in \eqref{eq:bounds}  reads $\langle V_{\rm up}^{\rm w}\rangle^{1/4}>m_{3/2}$.  
  For the extremely weakly warped throat,   the bulk KK mass scale is typically the lowest tower mass scale.
  Since it satisfies   $m_{\rm KK}\sim \langle V_{\rm up}^{\rm ew}/m_{\rm Pl}^4\rangle^{1/3} m_{\rm Pl}$, the first condition in \eqref{eq:bounds} is equivalent to $\langle V_{\rm up}^{\rm ew}/m_{\rm Pl}^4\rangle^{1/3} m_{\rm Pl}>m_{3/2}$.
  In any case, combined with the second condition in \eqref{eq:bounds}, one finds that the uplift potential  $V_{\rm up}^{\rm w}$ is bounded from both above and below.

   To see the behavior of $m_{3/2}$ more concretely, we first note that given the  K\"ahler potential
  \dis{\frac{K}{m_{\rm Pl}^2}=-2\log {\cal V}_0-\log(-i(\tau-\overline{\tau}))-\log\Big(\frac{i}{\kappa_4^6}\int \Omega\wedge\overline{{\Omega}}\Big)-\log\Big(\frac{1}{\kappa_4^6}\int d^6y\sqrt{g_6}e^{-4A}\Big),}  
    the flux-induced   Gukov-Vafa-Witten (GVW) superpotential is written as \cite{Gukov:1999ya}
  \dis{W = \frac{g_s^{3/2}}{\sqrt{4\pi}}\frac{m_{\rm Pl}^6}{\ell_s^2}\Big(\frac{m_{\rm Pl}}{m_s}\Big)^3 \int\Omega\wedge G_3,\label{eq:GVWsec}}
 the coefficient of which is obtained by matching  the flux term in the Type IIB supergravity action with the form of the  F-term potential \cite{Conlon:2005ki} (see also \cite{Bento:2021nbb}), as reviewed in Appendix \ref{App:GVW}.
 Then $m_{3/2}$ is given by
 \dis{m_{3/2}&=e^{K/(2m_{\rm Pl}^2)}\frac{|W|}{m_{\rm Pl}^2}
 \\
 &=\frac{g_s^2}{2\sqrt{2\pi}}\Big(\frac{1}{m_{\rm Pl}^6 i\int\Omega\wedge \overline{\Omega} }\Big)^{1/2}\frac{m_{\rm Pl}}{{\cal V}_0}\Big(m_{\rm Pl}^3 m_s^2 \int\Omega\wedge G_3 + ({\rm non\mbox{-}perturbative~terms})\Big)
 \\
 &=\frac{g_s^2}{2\sqrt{2\pi}}\Big(\frac{1}{m_{s}^6 i\int\Omega\wedge \overline{\Omega} }\Big)^{1/2}\frac{m_{\rm Pl}}{{\cal V}_0}\Big(m_s^5 \int\Omega\wedge G_3 + \sum_i A_i e^{i a_i \rho_i}\Big),}
 where in the last line, we replace $m_{\rm Pl}$  multiplied to $\Omega$ by $m_s$, as the complex structure moduli in $\Omega$   are typically written in units of $\ell_s$, just like $\epsilon^2=\ell_s^3 z$, to give $i \int\Omega\wedge\overline{\Omega}/\ell_s^6 \sim {\cal O}(1)$   under the normalization $\int d^6y\sqrt{g_6}e^{-4A}=\ell_s^6$.
 Moreover, even though the origins are different, the coefficients of the non-perturbative effects are written in the same way as the coefficient of the GVW superpotential for convenience. 
For later use, we define the dimensionless part of the superpotential by
  \dis{\widehat{W}=m_s^5 \int\Omega\wedge G_3 + \sum_i A_i e^{i a_i \rho_i}.}
 Indeed, the size of the GVW superpotential can be tuned by adjusting the amount of the harmonic $(0,3)$-form in $G_3$, which eventually determines the size of $\Lambda_{\rm SUSY}$.
 Such a tuning can be easily analyzed by considering $\widehat{W}$.

 Any model concerning the warped deformed conifold also requires the stabilization of the conifold modulus $z$, the complex structure modulus  determining the size of the warped throat.
 The $z$ dependent part of the K\"ahler potential is written as
 \dis{\frac{K}{m_{\rm Pl}^2}=\frac{\ell_s^6}{\pi i \int\Omega\wedge\overline{\Omega}}
\Big[|z|^2\Big(\log\frac{\Lambda_0^3}{|z|}+1\Big)+\frac{9 c'(g_sM)^2}{(2\pi)^4{\cal V}_0^{2/3}}|z|^{2/3}\Big]+\cdots,}
 then the F-term potential for $z$ is given by (see Appendix \ref{App:conV} for a review)
 \dis{V_{\rm KS}(|z|)=\frac{g_s^4}{8 {\cal V}_0^2}\Big[\log\frac{\Lambda_0^3}{|z|}+\frac{c'(g_s M)^2}{(2\pi)^4{\cal V}_0^{2/3}|z|^{4/3}}\Big]^{-1}\Big(\frac{M}{2\pi}\log\frac{\Lambda_0^3}{|z|}-\frac{K}{g_s}\Big)^2m_{\rm Pl}^4.\label{eq:Vforz}}
 This shows that the conifold modulus is stabilized at 
 $|z|= \Lambda_0^3 {\rm exp}[-\frac{2\pi K}{g_s M}]$.  
 In addition, the mass of the conifold modulus in the absence of uplift  is given by
 \dis{m_z^2&=\frac{m_{\rm Pl}^2}{K_{z\overline{z}}}V_{z\overline{z}}\Big|_{|z|= \Lambda_0^3 e^{-\frac{2\pi K}{g_s M}}}
 \\
 &=\frac{\pi i \int\Omega\wedge\overline{\Omega}}{\ell_s^6}\Big[\log\frac{\Lambda_0^3}{|z|}+\frac{ c'(g_sM)^2}{(2\pi)^4{\cal V}^{2/3}|z|^{4/3}}\Big]^{-2}\frac{g_s^4}{8{\cal V}_0^2}\Big(\frac{M}{2\pi}\Big)^2 \frac{m_{\rm Pl}^2}{2|z|^2}. \label{eq:zmass}}

\subsection{Conifold modulus mass scale}
\label{sec:conmass} 

 \subsubsection{Strongly warped throat} 
 
 We first consider the mass of the conifold modulus in the strongly warped throat in the sense of \eqref{eq:sscond}. 
 Since the term in the square brackets in \eqref{eq:zmass} is dominated by the second warping term,   we obtain
 \dis{m_z=\frac{4 \pi^{7/2}}{c'}\Big(\frac{ i \int\Omega\wedge\overline{\Omega}}{\ell_s^6}\Big)^{1/2}\frac{|z|^{1/3}}{M{\cal V}_0^{1/3}}m_{\rm Pl}.}
As  noted  previously,   the typical length scale associated with $\Omega$ is given by $\ell_s$ so we expect that $i\int\Omega\wedge\overline{\Omega}/\ell_s^6$ to be ${\cal O}(1)$.
Then $m_z$  depends on the same combination $|z|^{1/3}/{\cal V}_0^{1/3}$ as the KK mass scale $m_{\rm KK}^{\rm w}$ given by \eqref{eq:KKsw}, obeying  the relation
\dis{m_z \sim (2\pi)^3\eta_{\rm UV}m_{\rm KK}^{\rm w}\sim \frac{(2\pi)^2}{(g_s M^2 p)^{1/4}} \langle V_{\rm up}^{\rm w}\rangle^{1/4}.}
 Since the throat is strongly warped, i.e., $\eta_{\rm UV} > 1$, the KK  mass scale $m_{\rm KK}^{\rm w}$ is typically lighter than $m_z$.
  This indicates that the EFT based on the four-dimensional supergravity description  can be invalidated by the KK modes lighter than the conifold modulus.
 As claimed in \cite{Blumenhagen:2019qcg}, the light KK modes may be compatible with  the four-dimensional description for the stabilization of $|z|$ if the species cutoff above which the gravitational coupling in the loop becomes strong by the large number of particle species \cite{Dvali:2007hz, Dvali:2007wp} is well adjusted such that  the light KK modes just rescale the K\"ahler metric $K_{z \overline{z}}$ through the one-loop correction.
  We note that whereas the scaling behavior $m_z \sim  (g_s M^2 p)^{-1/4} \langle V_{\rm up}^{\rm w}\rangle^{1/4}$ works only for nonzero $p$, i.e., it has a discontinuity between zero and nonzero $p$, the relation  $m_z \sim \eta_{\rm UV}m_{\rm KK}$ is irrelevant to the existence of the uplift potential.

  We can understand the origin of the relation $m_z\sim |z|^{1/3}/{\cal V}_0^{1/3}$ in the following way.
  When the throat is strongly warped, the K\"ahler metric $K_{z\overline{z}}$ is enhanced by the warp factor $e^{-4A}\simeq e^{-4A_0}/\sigma \sim |z|^{-4/3}{\cal V}_0^{-2/3}$.
  In the F-term potential, the factors $e^K$ and $(K_{z\overline{z}})^{-1}$ provide ${\cal V}_0^{-2}$ and $e^{4A}$ respectively, and the redefinition of $\sigma(x)$ for the canonical kinetic term gives $(K_{z\overline{z}})^{-1}\sim e^{4A}$ in addition.
  Combining  them, we have  ${\cal V}_0^{-2}(e^{4A})^2\sim |z|^{8/3}/{\cal V}_0^{2/3}$.
 Meanwhile, $D_zW$   contains $\log(\Lambda_0^3/|z|)$, which originates from the monodromy behavior of $z$ around $z\simeq 0$.
 Then $V_{z\overline{z}}$ is dominated by the term containing $|(D_zW)_z|^2 \sim 1/|z|^2$ at the minimum of the potential.
 Combining all ingredients together,  one finds that $m_z^2\sim (|z|^{1/3}/{\cal V}_0^{1/3})^2$, consistent with the explicit result.
 While this reflects the throat geometry significantly, we did not find   a simple argument for the connection to   $m_{\rm KK}$ or $V_{\rm up}^{1/4}$.

 Our discussion so far is based on the assumption that $p$, the number of $\overline{\rm D3}$-branes is not so large that the mass scales are not much modified by $V_{\rm up}$ and the backreaction of $\overline{\rm D3}$-branes is negligibly small.
 As pointed out in \cite{Bena:2018fqc}, the sum of potential terms depending on $z$, $V_{\rm KS}+V_{\rm up}^{\rm w}$ (see  \eqref{eq:upsw} for $V_{\rm up}^{\rm w}$ and \eqref{eq:Vforz} for $V_{\rm KS}$)  stabilizes $z$ at the corrected value,
  \dis{|z|=\Lambda_0^3{\rm exp}\Big[-\frac{2\pi K}{g_sM}-\frac34 \pm\sqrt{\frac{9}{16}-\frac{4\pi p}{g_s M^2}\frac{2^{1/3}c'}{I(0)}}\Big],\label{eq:|z|correct}}
from which   $p$ is restricted to be smaller than $g_sM^2$.
When this bound is violated, $z$ is stabilized at $0$, giving $\eta_{\rm UV} \to \infty$, which is incompatible with the compact internal volume.
 Whereas it was argued that such runaway behavior may not appear when we take the off-shell contributions, i.e., quantum fluctuations around the stabilized values of moduli, into account \cite{Lust:2022xoq}, it is true that the geometry and the potential are drastically changed by the backreaction of  $\overline{\rm D3}$-branes for large $p$.
 We also note that $V_{\rm KS}$ and $V_{\rm up}^{\rm w}$ have the similar structure: since both $e^{K/m_{\rm Pl}^2}$ and $m_s^4/m_{\rm Pl}^4$ are proportional to $1/{\cal V}_0^2$,   $e^{K/m_{\rm Pl}^2}(K_{z\overline{z}})^{-1} \sim {\cal V}_0^{-2}e^{4A}$ in $V_{\rm KS}$ and $m_s^4 e^{4A}$ in $V_{\rm up}^{\rm w}$ contain the common factor $|z|^{4/3}/{\cal V}_0^{4/3}$.
 But $V_{\rm KS}$ also contains the $z$ dependent factor $|D_z W|^2$, which plays the crucial role in stabilizing $|z|$ at nonzero value.

  \subsubsection{Weakly warped throat}
  
  When the condition \eqref{eq:wwcond} is satisfied, the logarithmic term in $K^{z\overline{z}}$ dominates over the warping term.
  In the extremely weakly warped throat,   the logarithmic term   loses its dominance for $|z|/\Lambda_0^3 \simeq 1$, but at the same time the localization of $\overline{\rm D3}$-branes at the tip of the throat is not guaranteed.
  Therefore, in both cases,   so far as the valid EFT is concerned, the logarithmic term is the leading term in $K^{z\overline{z}}$.
  From \eqref{eq:zmass}, the conifold modulus mass in these cases is given by
  \dis{m_z=\frac{1}{8\sqrt{\pi}}\Big(\frac{ i \int\Omega\wedge\overline{\Omega}}{\ell_s^6}\Big)^{1/2}\frac{g_s^2M}{{\cal V}_0} \frac{m_{\rm Pl}}{|z|\log\frac{\Lambda_0^3}{|z|}} \sim \frac{g_s^2M}{{\cal V}_0} \frac{m_{\rm Pl}}{|z|\eta_{\rm UV}}.}
  Now we can compare this with  the lowest mass scale in   the weak and extremely weak  warping case, respectively.
  For the weakly warped throat in the sense of \eqref{eq:wwcond},  the ratio $m_{\rm KK}^{\rm w}/m_z \sim |z|^{4/3}{\cal V}_0^{2/3}/(g_sM)^2$  (see \eqref{eq:KKsw}) is smaller than $1/(2\pi)^4$  but larger than $1/[(2\pi)^4\eta_{\rm UV}]$.
Since $m_z$ is lighter than $m_{\rm KK}^{\rm w}$, just like the strongly warped case, the stabilization of the conifold modulus in the four-dimensional supergravity description can be invalid unless the light KK modes under the species cutoff just contribute to the rescaling of $K_{\rm z\overline{z}}$.
For the extremely weakly warped throat, the ratio  $m_{\rm KK}^{\rm ew}/m_z \sim |z|^{2/3}{\cal V}_0^{1/3}/(g_sM)$  (see \eqref{eq:mKKew})   and $m_{\rm KK}/m_z \sim |z|{\cal V}_0^{1/3}\eta_{\rm UV}/(g_sM)$ (see \eqref{eq:bulkKK}) are similar to or larger than $1/(2\pi)^2$ and $|z|^{1/3}\eta_{\rm UV}/(2\pi)^2$, respectively,  so as well known, $m_z$ can be lighter than the KK mass scale,  which is trustworthy provided $|z|/\Lambda_0^3$ is not too close to $1$.

Before discussing the correction to the stabilized value of $|z|$ by $V_{\rm up}$,  we note that  if we are also interested in the subleading terms in  the derivatives of the potential with respect  to $z$,  neglecting the  warping term from the beginning can be misleading.
To see this, consider $dV_{\rm KS}/d\overline{z}$  without assuming the weak  warping, given by (see \eqref{eq:Vforz})
{\small
\dis{\frac{1}{m_{\rm Pl}^4}\frac{dV_{\rm KS}}{d\overline{z}}=\frac{g_s^4}{8{\cal V}_0^2}\Big[&\Big(\log\frac{\Lambda_0^3}{|z|}+\frac{ c'(g_sM)^2}{(2\pi)^4{\cal V}_0^{2/3}|z|^{4/3}}\Big)^{-2}\Big(\frac12+\frac23\frac{c'(g_sM)^2}{(2\pi)^4{\cal V}_0^{2/3}|z|^{4/3}}\Big)\frac{1}{\overline{z}}\Big(\frac{M}{2\pi}\log\frac{\Lambda_0^3}{|z|}-\frac{K}{g_s}\Big)^2
\\
&-\Big(\log\frac{\Lambda_0^3}{|z|}+\frac{c'(g_sM)^2}{(2\pi)^4{\cal V}^{2/3}|z|^{4/3}}\Big)^{-1}\frac{1}{\overline{z}}\Big(\frac{M}{2\pi}\Big)  \Big(\frac{M}{2\pi}\log\frac{\Lambda_0^3}{|z|}-\frac{K}{g_s}\Big)\Big].\label{eq:VKSder}}}
In the second parenthesis in the first line, $1/2$ and the term containing $(g_sM)^2/[{\cal V}_0^{2/3}|z|^{4/3}]$ come from the derivative of $\log(\Lambda_0^3/|z|)$ and  the warping term with respect to $\overline{z}$, respectively.
While the latter is larger than the former under the condition \eqref{eq:wwcond}, it cannot be written if we ignore the warping term before taking derivative.
In order to compare the term in the first line with that in the second line,   let us impose the weak warping condition and ignore $1/2$ in the second parenthesis in the first line.
Since the warping term is subleading compared to $\eta_{\rm UV}= \log({\Lambda_0^3}/{|z|})$ in the last line, we obtain
{\small
\dis{\frac{1}{m_{\rm Pl}^4}\frac{dV_{\rm KS}}{d\overline{z}}\simeq \frac{g_s^4}{8{\cal V}_0^2}&\frac{1}{\eta_{\rm UV}}\Big(\frac{M}{2\pi}\Big)^2
 \Big(\log\frac{\Lambda_0^3}{|z|}-\frac{2\pi}{g_s}\frac{K}{M}\Big)\frac{1}{\overline{z}}
\Big[\Big(\frac23\frac{c'(g_sM)^2}{(2\pi)^4 \eta_{\rm UV}{\cal V}_0^{2/3}|z|^{4/3}}\Big)\Big(\log\frac{\Lambda_0^3}{|z|}-\frac{2\pi}{g_s}\frac{K}{M}\Big)-1\Big].}
}
In the square brackets, the first term comes from the dominant term of the first line ($1/2$ in the second parenthesis in the first line is ignored) and the second term comes from the second line in \eqref{eq:VKSder}, respectively.
This evidently shows that the term in the second line in \eqref{eq:VKSder} is most dominant : the term $\log\frac{\Lambda_0^3}{|z|}-\frac{2\pi}{g_s}\frac{K}{M}$ quickly approaches zero around the minimum and the coefficient is smaller then one.

In any case, the dominant term in $d(V_{\rm KS}+V_{\rm up})/d\overline{z}$ is written as
\dis{\frac{1}{m_{\rm Pl}^4}\frac{d(V_{\rm KS}+V_{\rm up})}{d\overline{z}}\simeq -\frac{g_s^4}{8{\cal V}_0^2}&\frac{1}{\eta_{\rm UV}}\Big(\frac{M}{2\pi}\Big)^2
 \Big(\log\frac{\Lambda_0^3}{|z|}-\frac{2\pi}{g_s}\frac{K}{M}\Big)\frac{1}{\overline{z}}+\frac{2^{4/3}\pi^3}{I(0)}\frac{g_s p}{M^2}\frac{1}{{\cal V}_0^{4/3}}\frac23\frac{z^{2/3}}{\overline{z}^{1/3}},}
and the stabilized value of $|z|$ is corrected to satisfy  $d(V_{\rm KS}+V_{\rm up})/d\overline{z}=0$.
It is convenient to parametrize the correction to $|z|$ by the shift   in $\eta_{\rm UV}$, the exponent of $|z|$, which will be denoted by $\varepsilon$ :
\dis{\log\frac{\Lambda_0^3}{|z|}=\frac{2\pi}{g_s}\frac{K}{M}+\varepsilon.\label{eq:varepsilon}}
Then we obtain
\dis{\varepsilon=\frac{8\pi}{3}\frac{2^{1/3}}{I(0)}\frac{p}{g_s M^2}\Big[\frac{(2\pi)^4}{(g_sM)^2}\eta_{\rm UV}|z|^{4/3}{\cal V}_0^{2/3}\Big].}
From \eqref{eq:wwcond}, one finds that  for the weakly warped throat, $\varepsilon$ lies in the range \dis{\frac{8\pi}{3}\frac{2^{1/3}}{I(0)}\frac{p}{g_s M^2} <\varepsilon < \frac{8\pi}{3}\frac{2^{1/3}}{I(0)}\frac{p}{g_s M^2}\eta_{\rm UV}.}
From the lower bound, we can say that the correction to the stabilized value of $|z|$ in the presence of the uplift is controllable  provided $p<g_sM^2/\eta_{\rm UV}$, similar to the bound on $p$ in the strongly warped throat.
For the extremely weakly warped throat,   the EFT is reliable only for  $\eta_{\rm UV}$   not too close to $0$.
Taking derivative of the second term in \eqref{eq:ewuplift} with respect to $\overline{z}$, we obtain
\dis{\frac{1}{m_{\rm Pl}^4}\frac{d(V_{\rm KS}+V_{\rm up})}{d\overline{z}}\simeq -\frac{g_s^4}{8{\cal V}_0^2}&\frac{1}{\eta_{\rm UV}}\Big(\frac{M}{2\pi}\Big)^2
 \Big(\log\frac{\Lambda_0^3}{|z|}-\frac{2\pi}{g_s}\frac{K}{M}\Big)\frac{1}{\overline{z}}
 +\frac{g_s^3}{4\pi}\frac{p}{{\cal V}_0^2}\frac{2^{2/3}I(0)(g_sM)^2}{(2\pi)^4|z|^{4/3}{\cal V}_0^{2/3}} \frac23\frac{1}{\overline{z}}.}
 Then $\varepsilon$ defined by \eqref{eq:varepsilon} is estimated as
 \dis{\varepsilon \simeq \frac{16\pi}{3}\frac{p\eta_{\rm UV}}{g_sM^2}\Big(\frac{2^{2/3}I(0)(g_sM)^2}{(2\pi)^4|z|^{4/3}{\cal V}_0^{2/3}}\Big).}
 The term in the parenthesis in the RHS is similar to or smaller than $1$ and also  $\eta_{\rm UV}$ for $\eta_{\rm UV}<1$, so  the value of $\varepsilon$ smaller than $1$ is allowed provided $p<(g_sM^2)/\eta_{\rm UV}$.

\subsection{Gravitino mass}
\label{sec:m32}

    We now move onto another  EFT validity condition $m_{3/2} < \Delta m$.
  When the throat is warped satisfying $|z|^{2/3}{\cal V}_0^{1/3} \ll (g_sM)/(2\pi)^2$, $m_{\rm KK}^{\rm w}$ is typically the lowest tower scale, then  this condition can be rewritten as
  \dis{\frac{g_s^2}{2\sqrt{2\pi}}\frac{1}{m_s^3\big(i\int\Omega\wedge \overline{\Omega}\big)^{1/2}}\frac{m_{\rm Pl}}{{\cal V}_0}\widehat{W} < \frac{2^{1/2}3^{1/6} \pi^{3/2}}{I(0)^{1/2}} \frac{|z|^{1/3}}{M \eta_{\rm UV} {\cal V}_0^{1/3}}m_{\rm Pl}.}
 Assuming $\int\Omega\wedge\overline{\Omega}/\ell_s^6 \sim {\cal O}(1)$, one finds that the fluxes and non-perturbative effects must be tuned such that $\widehat{W}$ satisfies at least
 \dis{\widehat{W}<\frac{1}{g_s^2 M \eta_{\rm UV}}|z|^{1/3}{\cal V}_0^{2/3}.\label{eq:m32mkk}}
 The upper bound on $\widehat{W}$   can be understood as follows. 
 The factor multiplied to $\widehat{W}$ in $m_{3/2}$ comes from $e^{K/(2m_{\rm Pl}^2)}$ which is proportional to $1/{\cal V}_0$, just like  $m_s^2 \propto 1/{\cal V}_0$.
 Moreover, while $e^{K/(2m_{\rm Pl}^2)}$ contains $g_s^{1/2}$, the coefficient of the GVW superpotential contains $g^{3/2}$ as can be found in \eqref{eq:GVWsec} (see also the last expression in \eqref{eq:GVW}), so $m_{3/2}$ is  proportional to $g_s^2$.
 Since  $m_s^2$ is also proportional to $g_s^2$, $m_{3/2}$ can be estimated as $m_{3/2}\sim (m_s^2/m_{\rm Pl})\widehat{W}$.
 Then the bound $m_{3/2}<m_{\rm KK}^{\rm w}$ becomes \eqref{eq:m32mkk}.

 We note that the bound $m_{3/2} < m_{\rm KK}^{\rm w}$  discussed above is just a minimum requirement, and depending on the model, we need to investigate if other moduli under consideration are still lighter than $m_{\rm KK}^{\rm w}$. 
  For the complex structure moduli, as we have seen, the conifold modulus is typically heavier than $m_{\rm KK}^{\rm w}$, which may invalidate  the model.
  \footnote{The stabilization of the axio-dilaton $\tau$ depends on the complex structure moduli and fluxes in the model.
  If the low energy effective superpotential is simply linear in $\tau$, the $\tau$ mass can be light or even destabilize the vacuum through the mixing with the volume modulus   \cite{Choi:2004sx, Seo:2021kyi}.
  }
  The mass scales of the K\"ahler moduli are also model dependent.
In the large volume scenario, the non-perturbative effect is dominated by that of the small cycle modulus, stabilizing the overall volume modulus at the large value (exponential in the small cycle modulus).
The overall volume modulus contributes to the potential through the K\"ahler potential only, hence each term in the potential shows the power-law dependence on  the overall volume modulus.
Then the moduli masses are similar to or even much lighter than $m_{3/2}$ \cite{Conlon:2005ki}.  
Meanwhile, in the KKLT scenario where  the single volume modulus $\sigma$ is taken into account, $V_{\sigma\sigma}\sim m_{3/2}^2$ but the normalization for the canonical kinetic term introduces a factor $(K_{\sigma\sigma})^{-1}\sim\sigma^2$ in addition, so the volume modulus mass is enhanced by $\sigma$, i.e., $m_\sigma \sim \sigma m_{3/2} \sim {\cal V}_0^{2/3} m_{3/2}$ \cite{Linde:2011ja}.
\footnote{We may understand why such an enhancement does not arise in the large volume scenario in the following toy example.
When we have a potential term $V\sim e^{-a_4 \tau_4}/\tau_5^{3/2}$ where $\tau_4$ is a small cycle modulus and $\tau_5$ is the overall volume modulus, $V_{\tau_5\tau_5}\sim e^{-a_4 \tau_4}/\tau_5^{2+3/2}$, but multiplying this by $(K_{\tau_5\tau_5})^{-1}\sim\tau_5^2$ cancels  $\tau_5^{-2}$ in   $V_{\tau_5\tau_5}$ again, resulting in $m_{\tau_5}^2\sim e^{-a_4 \tau_4}/\tau_5^{3/2}$, showing the same power-law dependence on $\tau_5$ as $V$.
This can be contrasted with the KKLT-type potential $V\sim e^{-\sigma}/\sigma$, in which $V_{\sigma\sigma}$ is dominated not by $\sim e^{-\sigma}/\sigma^{1+2}$ but by $V\sim e^{-\sigma}/\sigma$ as the derivative with respect to $\sigma$ can be taken on the exponential term as well.
Then  $m_\sigma^2$ can be enhanced by multiplying $(K_{\sigma\sigma})^{-1}\sim\sigma^2$. 
}
Requiring $m_\sigma <m_{\rm KK}^{\rm w}$, the bound on $\widehat{W}$ is much more constrained as
 \dis{\widehat{W}<\frac{1}{g_s^2 M \eta_{\rm UV}}|z|^{1/3}.\label{eq:KKLTWub}}

In the extremely weakly warped throat, the lowest tower mass scale is given by the bulk KK mass scale, so the condition can be written as
\dis{\frac{g_s^2}{2\sqrt{2\pi}}\frac{1}{m_s^3\big(i\int\Omega\wedge \overline{\Omega}\big)^{1/2}}\frac{m_{\rm Pl}}{{\cal V}_0}\widehat{W} < \sqrt{\pi}\frac{g_s}{{\cal V}_0^{2/3}}m_{\rm Pl}.}
From this we obtain
\dis{\widehat{W} <\frac{{\cal V}_0^{1/3}}{g_s},\label{eq:cond32ew}}
where the upper bound is nothing more than $m_{\rm Pl}/m_{\rm W}$, for the same reason as the bound on $\widehat{W}$ in the previous case.
 In the KKLT scenario, we can impose the condition $m_\sigma \sim {\cal V}_0^{2/3}m_{3/2} < m_{\rm KK}$ in addition, which provides the stronger bound $\widehat{W} <1/(g_s{\cal V}_0^{1/3})$.

\subsection{Preventing runaway}
\label{sec:run}
 
The condition $V_{\rm up}\lesssim {\cal O}(10)\times |V_{\rm AdS}|$ with $|V_{\rm AdS}| \leq \Lambda_{\rm SUSY}=3 m_{\rm Pl}^2 m_{3/2}^2$ is imposed to prevent the potential including the uplift from exhibiting the runaway behavior.
 We first consider the case in which the throat is warped satisfying $|z|^{2/3}{\cal V}_0^{1/3} \ll (g_sM)/(2\pi)^2$.
In the large volume scenario, SUSY is broken by F-term as well as $V_{\rm up}$, so $| V_{\rm AdS}|\sim \Lambda_{\rm SUSY} (\log {\cal V}_0/{\cal V}_0)$   is smaller than $\Lambda_{\rm SUSY}$.
Then   the condition reads
\dis{ \frac{2^{4/3}\pi^3}{I(0)}\frac{g_s p}{M^2}\frac{|z|^{4/3}}{{\cal V}_0^{4/3}} < {\cal O}(10)\times\frac{3 g_s^4}{8\pi} \frac{1}{m_s^6i\int\Omega\wedge\overline{\Omega}}\frac{\log{\cal V}_0}{{\cal V}_0^3}|\widehat{W}|^2,}
which is simplified to
\dis{|\widehat{W}|^2> {\cal O}(10^{-1})\times\frac{p}{g_s(g_sM)^2}|z|^{4/3}\frac{{\cal V}_0^{5/3}}{\log{\cal V}_0},}
where $\int\Omega\wedge\overline{\Omega}/\ell_s^6 \sim {\cal O}(1)$ is assumed.
Combined with the condition $m_{3/2}<m_{\rm KK}^{\rm w}$ given by \eqref{eq:m32mkk}, which is equivalent to the condition that the K\"ahler moduli are lighter than $m_{\rm KK}^{\rm w}$,  one obtains the  bound on $|\widehat{W}|^2$ given by
\dis{{\cal O}(10^{-1})\times\frac{p}{g_s(g_sM)^2}|z|^{4/3}\frac{{\cal V}_0^{5/3}}{\log{\cal V}_0} < |\widehat{W}|^2<\frac{1}{g_s^4 M^2 \eta_{\rm UV}^2}|z|^{2/3}{\cal V}_0^{4/3}.}
Comparing the lower and the upper bound, one finds that $|\widehat{W}|^2$ can exist only if
\dis{{\cal O}(10^{-1})\times g_s p\eta_{\rm UV}^2 \frac{|z|^{2/3}{\cal V}_0^{1/3}}{\log{\cal V}_0}<1}
is satisfied.
This inequality provides a bound on the number of $\overline{\rm D3}$-branes $p$, which can be taken into account in addition to $p< g_sM^2$ obtained from \eqref{eq:|z|correct}.
In fact, since $|z|^{2/3} {\cal V}_0^{1/3}\ll (g_sM)/(2\pi^2)$ is satisfied,   the LHS of the inequality is   smaller than $p(g_s^2 M)\eta_{\rm UV}^2 /[(2\pi)^2\log{\cal V}_0]$ times a small constant, say, of ${\cal O}(10^{-1})$.
  
In the KKLT scenario, the F-term potential is stabilized at the supersymmetric AdS minimum, so the runaway is prevented when $V_{\rm up} \lesssim {\cal O}(10)\times\Lambda_{\rm SUSY}$, which reads
\dis{\frac{2^{4/3}\pi^3}{I(0)}\frac{g_s p}{M^2}\frac{|z|^{4/3}}{{\cal V}_0^{4/3}} < {\cal O}(10)\times\frac{3 g_s^4}{8\pi} \frac{1}{m_s^6\int\Omega\wedge\overline{\Omega}}\frac{|\widehat{W}|^2}{{\cal V}_0^2},}
 or equivalently,
 \dis{|\widehat{W}|^2>{\cal O}(10^{-1})\times\frac{p}{g_s(g_sM)^2}|z|^{4/3}{\cal V}_0^{2/3}.}
 We note that the RHS of the inequality is smaller than $p/[(2\pi)^2 g_s]$ by the   condition $|z|^{2/3} {\cal V}_0^{1/3}\ll (g_sM)/(2\pi)^2$. 
 Combining this with \eqref{eq:KKLTWub}, one obtains  
 \dis{{\cal O}(10^{-1})\times\frac{p}{g_s(g_sM)^2}|z|^{4/3}{\cal V}_0^{2/3} <|\widehat{W}|^2 <\frac{1}{g_s^2(g_s M)^2 \eta_{\rm UV}^2}|z|^{2/3}, }
 which is valid only if 
 \dis{{\cal O}(10^{-1})\times g_s p\eta_{\rm UV}^2 |z|^{2/3}{\cal V}_0^{2/3}<1}
 is satisfied.
 This inequality can  be interpreted as a bound on $p$.

 We now consider the extremely weakly warped throat.
 The condition on the large volume scenario is given by 
 \dis{ \frac{g_s^3}{4\pi}\frac{p}{{\cal V}_0^2} < {\cal O}(10)\times\frac{3 g_s^4}{8\pi} \frac{1}{m_s^6i\int\Omega\wedge\overline{\Omega}}\frac{\log{\cal V}_0}{{\cal V}_0^3}|\widehat{W}|^2,}
 or equivalently, 
 \dis{|\widehat{W}|^2> {\cal O}(10^{-1})\times\frac{p}{g_s}\frac{{\cal V}_0}{ \log{\cal V}_0}.}
 Combined with the condition $m_{3/2}<m_{\rm KK}$ given by \eqref{eq:cond32ew}, one obtains 
 \dis{{\cal O}(10^{-1})\times\frac{p}{g_s}\frac{{\cal V}_0}{ \log{\cal V}_0} < |\widehat{W}|^2 < \frac{{\cal V}_0^{2/3}}{g_s^2},} 
which makes sense provided
\dis{p<{\cal O}(10)\times\frac{1}{g_s}\frac{\log{\cal V}_0}{{\cal V}_0^{1/3}}.}
Meanwhile, for the KKLT scenario, the condition $V_{\rm up} \lesssim {\cal O}(10)\times\Lambda_{\rm SUSY}$ is written as
\dis{\frac{g_s^3}{4\pi}\frac{p}{{\cal V}_0^2}  < {\cal O}(10)\times\frac{3 g_s^4}{8\pi} \frac{1}{m_s^6\int\Omega\wedge\overline{\Omega}}\frac{|\widehat{W}|^2}{{\cal V}_0^2},}
 which becomes the volume-independent condition
 \dis{|\widehat{W}|^2>{\cal O}(10^{-1})\times\frac{p}{g_s}.}
 Combined with the bound $m_\sigma <m_{\rm KK}$ given by $\widehat{W} <1/(g_s{\cal V}_0^{1/3})$, we obtain
 \dis{{\cal O}(10^{-1})\times\frac{p}{g_s} < |\widehat{W}|^2 < \frac{1}{g_s^2{\cal V}_0^{2/3}},}
 which is valid provided $p<{\cal O}(10)\times[1/(g_s {\cal V}_0^{2/3})]$.

 \section{Conclusions}
\label{sec:conclusion}

 In this article, we investigate the connection between the uplift and the distance conjecture by  considering the concrete model, the warped deformed conifold embedded into Type IIB flux compactification with the uplift produced by $\overline{\rm D3}$-branes at the tip of the throat.
 Whereas  the various mass scales associated with  towers of states can be found, it turns out that the lowest tower mass scale  obeys the scaling behavior with respect to $V_{\rm up}$, which is meaningful only if the number of  $\overline{\rm D3}$-branes is nonzero.  
Then in the $V_{\rm up} \to 0$ limit, the EFT becomes invalid by the descent of a tower of states   from UV, as the distance conjecture predicts.  
 Since too large $V_{\rm up}$ also is not allowed in the EFT due to the sizeable backreaction and the possible runaway behavior of the moduli potential, the size of $V_{\rm up}$ consistent with the EFT is bounded from both above and below.
 In the simple model like the KKLT or the large volume scenario in which the non-perturbative effect is dominated by the single term, this bound can be rewritten as the bound on the size of the superpotential.
  
  The bound we found can be more restricted depending on the details of the model.
  For instance, if mass of the volume modulus becomes very heavy due to the tuning between more than two non-perturbative terms in the superpotential, the tower mass scale obeying the scaling behavior with respect to  $V_{\rm up}$  is required to be   heavier than the volume modulus mass.
 At the same time, the lower bound on $V_{\rm up}$ in this case is no longer $|V_{\rm AdS}|$ but the sum of  $|V_{\rm AdS}|$ and the height of $V_{\rm AdS}$  at the local maximum, so it cannot be rewritten as a bound on the superpotential in a simple manner.
 
 Another issue which is not addressed  in this article is that, whereas we simply assume $e^{-4A} \simeq 1$ outside the throat, the too large value of $g_s MK$ compared to ${\cal V}_0$ can result in the existence of the singular points in the bulk region at which $e^{-4A}$ becomes zero or even negative  \cite{Gao:2020xqh}.
 This leads to the serious control issue of the KKLT scenario with almost vanishing $\Lambda$.
 We may avoid this problem   in the large volume scenario or  the moduli  stabilization in the (A)dS vacuum, but the constraints on $V_{\rm up}$ and the connection    to the distance conjecture is the subject of the future study.
 
 On the other hand, when more than two throats are related homologically, the length of each throat is shorter than $\log(\Lambda_0^3/|z|)$ as the $H_3$-flux accumulated in each throat is smaller than $K$ \cite{Hebecker:2018yxs, Carta:2021uwv}.
 Moreover, when   $\overline{\rm D3}$-branes are located at the tip of only one of throats, the corresponding throat is no longer equivalent to other throats \cite{Seo:2021zyc}.
 Then through the brane/flux annihilation, the uplift potential as well as the throat geometry changes until SUSY is restored.
 In these cases, we may need to revisit the criterion distinguishing the strong warping from  the weak warping.
 Moreover, the hierarchies between mass scales are not simple as what we discussed in this article.
 We expect that such nontrivial model dependent features are helpful to understand the naturalness criterion on the string models, especially those realizing the tiny cosmological constant as we observe it, in light of the  distance conjecture.


%


\appendix

\section{Review on Klebanov-Strassler throat}
\label{App:KSreview}

In this appendix we summarize the features of the background geometry described by the Klebanov-Strassler   throat \cite{Klebanov:2000hb}, a noncompact, asymptotically conical solution of Type IIB supergravity supported by the fluxes.
The metric of the Klebanov-Strassler   throat is given by
{\small
\dis{& ds_{\rm con}^2=\frac{\epsilon^{4/3}K(\eta)}{2}\Big[\frac{1}{3K(\eta)^3}(d\eta^2+(g^5)^2)+\sinh^2\Big(\frac{\eta}{2}\Big)((g^1)^2+(g^2)^2)+\cosh^2\Big(\frac{\eta}{2}\Big)((g^3)^2+(g^4)^2)\Big],}}
where 
\dis{K(\eta)=\frac{(\sinh(2\eta)-2\eta)^{1/3}}{2^{1/3}\sinh \eta}.}
Here $\epsilon$ parametrizes the deformation of the tip of the throat, i.e., smoothing out the $S^3$ singularity of the  $T^{1,1}\sim S^3\times S^2$ base described by $\sum_{A=1}^4 w_A^2=\epsilon^2$ with $w_A \in \mathbb{C}$ and the basis of 1-forms $\{g^i\}$ ($i=1,\cdots,4$).
The deformation is also parametrized by $z \equiv \epsilon^2/\ell_s^3$, which is dimensionless and interpreted as the stabilized value of the conifold modulus, the complex structure modulus determining the size of $\epsilon$.
Whereas the $S^3$ subspace is referred to as the A-cycle, the $S^2\times \mathbb{R}$ subspace in which $\mathbb{R}$ is parametrizerd by $\eta$ is called the B-cycle.
Here   $\eta$ extends over  $[0, \eta_{\rm UV}]$, where $\eta_{\rm UV}$ is the coordinate at which the throat is glued to the compact bulk.
 Then the features of the geometry near and far   from  the tip of the throat can be found by taking the limits $\eta \ll 1$ and $\eta \gg 1$, respectively.

\subsection{Geometry near the tip ($\eta \ll 1$ limit)}

In the $\eta \ll 1$ limit, the metric is approximated as
\dis{ds_{\rm con}^2\simeq \frac{\epsilon^{4/3}}{4}\Big(\frac23\Big)^{1/3} \Big[d\eta^2+\frac12((g^1)^2+(g^2)^2)+2\Big(\frac12 (g^5)^2+(g^3)^2+(g^4)^2\Big)\Big],}
where $(g^1)^2+(g^2)^2$ and $\frac12 (g^5)^2+(g^3)^2+(g^4)^2$ describes $S^2$ and $S^3$ of radius $\sqrt2$, respectively.
With the appropriate choice of the coordinates we may rewrite it as (see, e.g., \cite{Nguyen:2019syc} and references therein)
\dis{ds_{\rm con}^2\simeq \frac{\epsilon^{4/3}}{4}\Big(\frac23\Big)^{1/3}\Big[  d\eta^2+\eta^2 (d\widetilde{\omega}^2+\sin\widetilde{\omega}d\widetilde{\varphi}^2)+4(d\psi^2+\sin^2\psi(d\omega^2+\sin^2\omega d\varphi^2)\Big].\label{eq:KStipmetric}}
On the other hand, the warp factor $e^{-4A}=1+ e^{-4A_0}/\sigma(x)$ where
\dis{&e^{-4A_0(y)}\simeq 2^{2/3}\frac{(\alpha' g_sM)^2}{\epsilon^{8/3}}I(\eta)
\\
& I(\eta)=\int_\eta^\infty dx \frac{x\coth x-1}{\sinh^2 x}(\sinh(2x)-2x)^{1/3},}
is dominated by $e^{-4A_0}/\sigma$ when   $z = \epsilon^2/\ell_s^3$ is so small that $|z|^{2/3} {\cal V}_0^{1/3}\ll (g_sM)/(2\pi)^2$ is satisfied. 
In this case, the factor $\epsilon^{4/3}\sigma^{1/2}$ in the denominator of $e^{-2A}\simeq e^{-2A_0}/\sigma^{1/2}$ is cancelled by the prefactor $\sigma^{1/2}$ in $G_{mn}=e^{-2A}\sigma^{1/2}g_{mn}$ (see  \eqref{eq:metric}) and $\epsilon^{4/3}$ in $g_{mn}$ (more precisely, $ds^2_{\rm con}$), respectively.
Then the ten-dimensional metric near $\eta\simeq 0$ is written as
\dis{ds^2&\simeq \Big(\frac23\Big)^{1/2}\frac{1}{b_0^2}\frac{\epsilon^{4/3}}{(\alpha' g_sM)}\frac{\langle\sigma^{3/2}\rangle}{\sigma(x)}g_{\mu\nu}dx^\mu dx^\nu
\\
&+\frac{b_0^2}{4}(\alpha' g_sM)\Big[  d\eta^2+\eta^2 (d\widetilde{\omega}^2+\sin\widetilde{\omega}d\widetilde{\varphi}^2)+4(d\psi^2+\sin^2\psi(d\omega^2+\sin^2\omega d\varphi^2)\Big],\label{eq:sw10Dmetric}}
where $b_0^2=(4/3)^{1/3}I(0)^{1/2}$ with $I(0)\simeq 0.71805$.
While $\epsilon$ does not appear in the six-dimensional internal space metric, as we will see in the discussion on the $\eta \gg 1$ limit, it determines the length of the throat.

\subsection{Geometry far  from the tip ($\eta \gg 1$ limit)}

To see the $\eta \gg 1$ limit of the geometry, it is convenient to define the `radial coordinate'
\dis{r=\frac{3^{1/2}}{2^{5/6}}\epsilon^{2/3}e^{\eta/3},\label{eq:rdef}}
in terms of which the metric is approximated as
\dis{ds_{\rm con}^2 \simeq dr^2+r^2\Big(\frac19(g^5)^2+\frac16\sum_{i=1}^4 (g^i)^2\Big) = dr^2+r^2 ds_{T^{1,1}}^2.}
 Whereas  the warp factor $e^{-4A}=1+(e^{-4A_0}/\sigma)$ becomes $1$ outside the throat, when $e^{-4A_0}/\sigma > 1$ is satisfied in  the throat region,  $e^{-4A_0}$ is approximated as 
\dis{e^{-4A_0(y)}\simeq \frac{L^4}{r^4}\Big[1+\frac{3g_sM}{2\pi K}\log\Big(\frac{r}{r_{\rm UV}}\Big)+\frac{3g_sM}{8\pi K}\Big],\quad
L^4=\frac{27\pi}{4}g_s MK{\alpha'}^2,\label{eq:e4Asw}} 
where $r_{\rm UV}=r(\eta=\eta_{\rm UV})$.
Denoting $r(\eta=0)$ by $\eta_{\rm IR}$, the sum of first two terms, which can be rewritten as $[(3g_sM)/(2\pi K)]\log[r/r_{\rm IR}]$, represents the accumulation of the $H_3$-flux along the B-cycle,  satisfying $K=[(3g_sM)/(2\pi)]\log[r_{\rm UV}/r_{\rm IR}]=[(g_sM)/(2\pi)]\eta_{\rm UV}$, where in the last equality \eqref{eq:rdef} is used.
Using \eqref{eq:rdef} again,  one finds that $z=\epsilon^2/\ell_s^3$ must be stabilized at $\Lambda_0^3 {\rm exp}[-\frac{2\pi K}{g_s M}]$, with $\Lambda_0=(2^{5/6}/3^{1/2}) (r_{\rm UV}/\ell_s) $ \cite{Giddings:2001yu}. 
Then we learn that $\eta_{\rm UV}$, or equivalently, $\frac{2\pi K}{g_s M}$ becomes larger as the warping gets stronger.

For the strongly warped throat,  the ten-dimensional metric near $r_{\rm UV}$ is close to
\dis{ds^2\simeq \Big(\frac{r}{R}\Big)^2\frac{\langle \sigma^{3/2}\rangle}{\sigma(x)}g_{\mu\nu}dx^\mu dx^\nu+\Big(\frac{R}{r}\Big)^2(dr^2+r^2 ds_{T^{1,1}}^2),}
where $R^4=L^4(1+[(3g_sM)/(8\pi K)])$ and we used \eqref{eq:e2Omega} for $e^{2\Omega}$. 
After the stabilization of $\sigma$ to $\langle\sigma\rangle={\cal V}_0^{2/3}$ and the rescaling $r \to r/\langle\sigma^{1/4}\rangle=r/{\cal V}_0^{1/6}$, the geometry becomes AdS$_5\times$T$^{1,1}$.
We note that  for  $e^{-4A_0}/\sigma > 1$,   $r_{\rm UV}$ is restricted to be smaller than $R/\sigma^{1/4}$, or roughly $(g_s MK)^{1/4} \ell_s/{\cal V}_0^{1/6}$ (after rescaling, $r_{\rm UV}<R\simeq (g_s MK)^{1/4} \ell_s$).
Therefore, $R$ is interpreted as the radial size of the throat, which is required to be smaller than the overall volume size, i.e.,  $R<\sigma^{1/4}$, or $(g_s MK)^{1/4} < {\cal V}_0^{1/6}$ \cite{Carta:2019rhx}.

\subsection{Stabilization of  the conifold modulus}
\label{App:conV}

The size of $\epsilon^2=z\ell_S^2$ is determined by the stabilization of the  conifold modulus.
The K\"ahler potential for $z$ is studied in \cite{Douglas:2007tu} (see also \cite{Lust:2022xoq} and Appendix A of \cite{Bento:2021nbb}), which we will briefly sketch here.
Using the fact that the warp factor $e^{-4A}\simeq 1$ in the bulk and denoting $-\log (\frac{i}{\kappa_4^6}\int_{\rm bulk}\Omega\wedge\overline{\Omega} )$ by $K_{\rm cs}^{\rm bulk}$, the K\"ahler potential for the complex structure moduli is written as
\dis{\frac{K_{\rm cs}}{m_{\rm Pl}^2}&=-\log\Big(\frac{i}{\kappa_4^6}\int e^{-4A}\Omega\wedge\overline{\Omega}\Big)=-\log \Big(\frac{i}{\kappa_4^6}\int_{\rm bulk}\Omega\wedge\overline{\Omega} + \frac{i}{\kappa_4^6}\int_{\rm conifold}e^{-4A}\Omega\wedge\overline{\Omega}\Big)
\\
&\simeq K_{\rm cs}^{\rm bulk}+e^{K_{\rm cs}^{\rm bulk}} \frac{i}{\kappa_4^6}\int_{\rm conifold}e^{-4A}\Omega\wedge\overline{\Omega},}
from which the K\"ahler metric for the complex structure moduli, represented by the   harmonic $(2,1)$-form $\chi_a$, is given by $K_{a{\overline{b}}}=(i\int e^{-4A}\chi_a\wedge \overline{\chi}_{\overline b})/(i\int e^{-4A}\Omega \wedge \overline{\Omega} )$.
For the conifold modulus $S\equiv \epsilon^2=\ell_s^3 z$ which is localized in the throat, the numerator of $G_{a{\overline{b}}}$ is dominated by the throat part, whereas the denominator is still dominated by the bulk part, i.e.,
\dis{K_{S\overline{S}}=\frac{i\int_{\rm conifold} e^{-4A}\chi_S\wedge \overline{\chi}_{\overline S}}{i\int_{\rm bulk} e^{-4A}\Omega \wedge \overline{\Omega}}=e^{K_{\rm cs}^{\rm bulk}} \frac{i}{\kappa_4^6}\int_{\rm conifold} e^{-4A}\chi_S\wedge \overline{\chi}_{\overline S},}
where
\dis{&\chi_S=g^3\wedge g^4\wedge g^5+d\big[F(\eta)(g^1\wedge g^3+g^2\wedge g^4)\big]-id\big[f(\eta)(g^1\wedge g^2)+k(\eta)(g^3\wedge g^4)\big],
\\
&F(\eta)=\frac{\sinh\eta-\eta}{2\sinh\eta},\quad
f(\eta)=\frac{\eta\coth \eta-1}{2\sinh \eta}(\coth \eta-1),\quad
k(\eta)=\frac{\eta\coth \eta-1}{2\sinh \eta}(\coth \eta+1).}
In two limits $\eta \ll 1$ and $\eta \gg 1$, three functions used to define $\chi_S$ behave as 
\dis{\eta\to 0\quad &:\quad F(0)=\frac{\eta^2}{12},\quad f(0)=\frac{\eta^3}{12},\quad k(0)=\frac{\eta}{3},
\\
\eta\to \infty\quad &:\quad F(\infty)=\frac12-\eta e^{-\eta}, \quad f(\infty)=\frac{\eta }{2},\quad k(\infty)=\frac{\eta}{2}.} 
Putting 
\dis{\chi_S\wedge \overline{\chi}_{\overline S}=-\frac{2i}{64\pi^3}d\eta\wedge\big(\prod_i g^i\big)\frac{d}{d\eta}[f(\eta)+F(\eta)(k(\eta)-f(\eta)]}
and using $\int \prod_i g^i=64\pi^3$, the K\"ahler metric is written as
\dis{\frac{K_{S\overline{S}}}{m_{\rm Pl}^2}&=\frac{2}{\pi i\int\Omega\wedge\overline{\Omega}}\int d \eta e^{-4A}\frac{d}{d\eta}[f(\eta)+F(\eta)(k(\eta)-f(\eta)]
\\
&=\frac{2}{\pi i\int\Omega\wedge\overline{\Omega}}\int d \eta \Big[\frac{d}{d\eta}[e^{-4A}(f+F(k-f)]-\frac{d e^{-4A}}{d\eta}[f+F(k-f)]\Big].}
The first integral can be evaluated by noting that $f+F(k-f)$ becomes $0$ for $\eta\to 0$ and $\frac{\eta_{\rm UV}}{2}=\frac34\log(\frac{2^{5/3}}{3})+\frac12\log(\frac{r_{\rm UV}^3}{S})$ (see \eqref{eq:rdef}) for $\eta\to \infty$ where $e^{-4A}\simeq 1$.
For the second integral, one can use the fact that
\dis{\frac{d e^{-4A}}{d\eta}=\frac{1}{\sigma(x)}\frac{d e^{-4A_0}}{d\eta}
=\frac{2^{2/3}(\alpha' g_sM)^2}{\sigma(x) S^{4/3}}
\frac{d I(\eta)}{d\eta}
=-4\frac{2^{2/3}(\alpha' g_sM)^2}{\sigma(x) S^{4/3}}\frac{f+F(k-f)}{(\sinh(2\eta)-2\eta)^{3/2}},}
to evaluate the integral numerically.
We note that while the first integral is dominated by the region $\eta\simeq \eta_{\rm UV}$ at which $e^{-4A}\simeq 1$, the second integral contains the variation of the warp factor, which is enhanced for  the sizeable  $e^{-4A_0}/\sigma$ in the throat region.
Then we have
\dis{\frac{K_{S\overline{S}}}{m_{\rm Pl}^2}=\frac{1}{\pi i\int\Omega\wedge\overline{\Omega}}\Big[\log\Big(\frac{r_0^3}{S}\Big)+c'\frac{(\alpha' g_SM)^2}{\sigma(x) S^{4/3}}\Big],\label{eq:Kforz}}
where $r_0=(2^{5/3}/3)^{1/2}r_{\rm UV}$ and $c'\simeq 1.18$.
This can be obtained from the K\"aher potential
\dis{\frac{K}{m_{\rm Pl}^2}=&-2\log{\cal V}_0 +\frac{\ell_s^6}{\pi \int\Omega\wedge\overline{\Omega}}
\Big[|z|^2\Big(\log\frac{\Lambda_0^3}{|z|}+1\Big)+\frac{9 c'(g_sM)^2}{(2\pi)^4{\cal V}^{2/3}}|z|^{2/3}\Big]
\\
\simeq & -3\log\Big({\cal V}_0^{2/3}-
\frac{3 c'(g_sM)^2 \ell_s^6}{(2\pi)^4\pi i\int\Omega\wedge\overline{\Omega} }|z|^{2/3}\Big)+
\frac{\ell_s^6}{\pi i\int\Omega\wedge\overline{\Omega}}
|z|^2\Big(\log\frac{\Lambda_0^3}{|z|}+1\Big)
,}
where $\Lambda_0=r_{\rm UV}/\ell_s$.

 Meanwhile, the flux induced GVW superpotential  can be explicitly written by introducing $(\alpha_I, \beta^I)$ ($I=0,\cdots, h^{2,1}$), the basis of the de Rham cohomology group $H^3(\mathbb{Z})$ and $(A^I, B_I)$, the Poincar\'e dual homology basis satisfying
 \dis{&A^I\cdot A^J=0=B_I\cdot B_J,\quad A^I\cdot B_J=\delta^I_J,
 \\
 &\int_{A_J}\alpha_I=-\int_{B_I}\beta^J=\int_{\rm CY_3}\alpha_I\wedge \beta^J=\delta_I^J,}
 such that the fluxes are quantized as
 \dis{F_3=\ell^2(M^I\alpha_I+M_I\beta^I),\quad H_3=\ell^2(K^I\alpha_I+K_I\beta^I),}
 which is constrained by the tadpole condition
 \dis{\frac{1}{\ell_s^2}\int_{\rm CY_3}F_3\wedge H_3=M^IK_I-M_IK^I=-N_{\rm D_3}+\frac12 N_{O_3}+\frac{\chi}{24},}
 where $\chi$ is the Euler characteristic of CY$_4$ in the F-theory compactification.
 Meanwhile, the holomorphic 3-form $\Omega$ is written as
 \dis{\Omega=Z^I \alpha_I-{\cal F}_I \beta^I,}
 where
\dis{Z^I=\int_{A^I}\Omega,\quad {\cal F}_I=\int_{B_I}\Omega.}
Then the complex structure moduli are identified with $t^a=Z^a/Z^0$ ($a=1,\cdots, h^{2.1}$) and prepotential $F(t^a)$ can be defined as ${\cal F}(Z^I)=(Z^0)^2F(t^a)$ with ${\cal F}_I=\partial_I{\cal F}$, from which the GVW superpotential (for the coefficient, see the first expression in \eqref{eq:GVW}) is written as
\footnote{In the same way, the K\"ahler potential for the complex structure moduli can be written as
\dis{K_{\rm cs}=-\log\Big(\int\Omega\wedge\overline{\Omega}\Big)=-\log(i|Z^0|^2[2(F-\overline{F})-(t^a-\overline{t}^a)(F_a+\overline{F}_a)]).}}
\dis{W=(g_s {\cal V}_0)^{1/2} m_{\rm Pl}^8\ell_s^2[(Z^IM_I+F_IM^I)-\tau(Z^IK_I+F_IK^I)].}
 In particular, in terms of the A-cycle($S^3$) and the B-cycle ($S^2\times\mathbb{R}$) of the throat satisfying
 \dis{\int_A F_3=\ell_s^2M,\quad\int_B H_3=-4\pi^2 K,\quad\int_A H_3=\int_B F^3=0,}
 i.e., $(M^S, M_S)=(M,0)$ and $(K^S, K_S)=(0,K)$, the conifold modulus and the corresponding prepotential are given by 
 \dis{S =\int_A\Omega=\ell_s^3 z,\quad {\cal F}_S=\int_B\Omega=\frac{S}{2\pi i}\Big[\log\Big(\frac{S}{r_0}\Big)-1\Big]+({\rm regular~terms}),}
 respectively.
Then the superpotential is written as
\dis{W=(g_s {\cal V}_0)^{1/2}m_{\rm Pl}^3\Big(\frac{m_{\rm Pl}}{m_s}\Big)^5\Big[M\frac{z}{2\pi i}\Big[\log\Big(\frac{S}{r_0}\Big)-1\Big]-\frac{i}{g_s}Kz+\cdots\Big].\label{eq:Wforz}}
 
Using \eqref{eq:Kforz} and \eqref{eq:Wforz}, together with the K\"ahler potential given by \eqref{eq:Kahlergen} (giving $e^{K/m_{\rm Pl}^2}={\cal V}_0^{-1/2}(g_s/2)(i m_{\rm Pl}^6\int\Omega\wedge\overline{\Omega})^{-1}(m_s/m_{\rm Pl})^6$) one finds  the F-term potential for $z$ given by \eqref{eq:Vforz}.

\section{Uplift potential in terms of NS5-brane}
\label{App:NS5}

In this appendix, we sketch how to obtain the  uplift potential in terms of NS5-brane, which is extended over the four-dimensional noncompact spacetime and wraps the $S^2$ subspace of the A-cycle ($S^3$ part of the throat).
From \eqref{eq:metric} and \eqref{eq:KStipmetric}, the induced metric on NS5-brane is written as
\dis{ds_{\rm NS5}^2=e^{2A(y)}e^{2\Omega(x)}\eta_{\mu\nu}dx^\mu dx^\nu+e^{-2A(y)}\sigma(x)^{1/2}\Big(\frac23\Big)^{1/3}\epsilon^{4/3}\sin^2\psi(d\omega^2+\sin^2\omega d\varphi^2).}
Meanwhile, the NS5-brane action is given by
\dis{-S_{\rm NS5}=\frac{T_5}{g_s^2}\int d^6\xi \sqrt{-g}\sqrt{{\rm det}(g_{2-{\rm cycle}}+g_s 2\pi \alpha'{\cal F}_2)} + T_5\int B_6,}
where $T_5=2\pi \ell_s^{-6}=T_3 /[(2\pi)^2\alpha']$ (the D$p$-brane tension is given by $T_p=(2\pi)^{-p}(\alpha')^{-\frac{p+1}{2}}=2\pi \ell_s^{-(p+1)}$) and $2\pi \alpha'{\cal F}=2\pi \alpha' F_2+C_2$ with
\dis{&2\pi \alpha' \int_{S^3}F_2=-4\pi^2 p,
\\
&C_2=M\big(\psi-\frac12 \sin 2\psi\big)\sin\omega d\omega \wedge d\varphi+\cdots.}
We also note that the imaginary self-duality $\star_6 G_3=i G_3$ gives $\star_6 H_3=-g_s F_3$.
Then we obtain
\dis{V_{\rm NS5}=&\frac{1}{g_s^2}\Big(\frac{T_3}{(2\pi)^2 \alpha'}\Big)e^{4A}e^{4\Omega}4\pi(
\alpha' g_sM)
\\
&\quad \times\Big[\sqrt{\frac{e^{-4A}\sigma(2/3)^{2/3}\epsilon^{8/3}}{(\alpha'g_sM)^2}\sin^4\psi+\Big(\frac{\pi p}{M}-\big(\psi-\frac12\sin 2\psi\big)\Big)^2}+\Big(\frac{\pi p}{M}-\big(\psi-\frac12\sin 2\psi\big)\Big)\Big]
\\
=&\frac{T_3}{g_s}\frac{M}{\pi}e^{4A}e^{4\Omega}
\\
&\quad
\Big[\sqrt{\frac{e^{-4A}\sigma(2/3)^{2/3}\epsilon^{8/3}}{(\alpha'g_sM)^2}\sin^4\psi+\Big(\frac{\pi p}{M}-\big(\psi-\frac12\sin 2\psi\big)\Big)^2}+\Big(\frac{\pi p}{M}-\big(\psi-\frac12\sin 2\psi\big)\Big)\Big].}
For $\psi=0$, the potential is reduced to the $\overline{\rm D3}$-brane uplift potential given by \eqref{eq:uplift}, as the term in the square brackets becomes $2\pi p/M$.
Indeed, when the condition $|z|^{2/3} {\cal V}_0^{1/3}\ll (g_sM)/(2\pi^2)$ is satisfied,  the coefficient of $\sin^4\psi$ in the square root is simplified to $b_0^4$ such that
\dis{V_{\rm NS5}=&\frac{2^{1/3}}{I(0)}\Big(\frac{g_s^3}{8\pi}\Big)m_{\rm Pl}^4 \frac{(2\pi)^4 |z|^{4/3}}{(g_sM)^2\sigma(x)^2}\frac{M}{2\pi}
\\
&\times\Big[\sqrt{b_0^4\sin^4\psi+\Big(\frac{\pi p}{M}-\big(\psi-\frac12\sin 2\psi\big)\Big)^2}+\Big(\frac{\pi p}{M}-\big(\psi-\frac12\sin 2\psi\big)\Big)\Big].}
The ${\cal O}(1)$ coefficient $2^{1/3}/I(0)\simeq 1.75$ is often denoted by $c''$.
We note that in contrast to the D$p$-brane action which is proportional to $g_s^{-1}$,  the NS5-brane action is proportional to $g_s^{-2}$, but since the ${\cal F}$-flux contribution is proportional to $g_s$, $V_{\rm NS5}$ can be reduced to $V_{\overline{\rm D3}}$ for $\psi=0$.

\section{Coefficient of Gukov-Vafa-Witten superpotential}
\label{App:GVW}

Here we sketch how to fix the coefficient of the GVW superpotential, following Appendix A of \cite{Conlon:2005ki} (see also \cite{Bento:2021nbb}).
We begin with the fact that when the fluxes are turned on, the  $|G_3|^2$ term in the Type IIB supergravity action \eqref{eq:TypeIIBaction} gives the potential for $G_3^{\rm IASD}$, the imaginary anti-self dual part of $G_3$ consisting of the harmonic $(3,0)$- and $(1,2)$-forms (for derivation, see, e.g., Appendix in \cite{Giddings:2001yu}).
This is interpreted as the F-term potential obtained from the GVW superpotential : 
\dis{V_{\rm flux}&=\frac{g_s}{2\kappa_{10}^2}\int \frac{G_3^{\rm IASD}\wedge\overline{G_3^{\rm IASD}}}{{\rm Im}\tau}
\\
&=\frac{g_s}{\kappa_{10}^2}\frac{i}{2 {\rm Im}\tau \int\Omega\wedge\overline{{\Omega}}} \sum_{a,b}(m_{\rm Pl}^2 K^{a\overline{b}})D_aW_{\rm GVW} D_b\overline{W}_{\rm GVW},}
where 
\dis{W_{\rm GVW}=\int \Omega\wedge G_3.}
Here the indices $a, b$ run over the complex structure moduli as well as  $\tau$ and the inverse of the K\"ahler metric $K^{a\overline{b}}$ is  obtained from the K\"ahler potential
\dis{\frac{K}{m_{\rm Pl}^2}=-2\log {\cal V}_0-\log(-i(\tau-\overline{\tau}))-\log\Big(\frac{i}{\kappa_4^6}\int \Omega\wedge\overline{{\Omega}}\Big)-\log\Big(\frac{1}{\kappa_4^6}\int d^6y\sqrt{g_6}e^{-4A}\Big),\label{eq:Kahlergen}}
which consists of   the K\"ahler potential for the overall volume (K\"ahler) modulus, axio-dilaton, complex structure moduli, and K\"ahler moduli other than the overall volume modulus.
We note that whereas the    last term is not taken into account in \cite{Conlon:2005ki}, it does not affect our discussion  so far as all the K\"ahler moduli other than the overall volume modulus are heavier than the energy scale of the EFT   (see \cite{DeWolfe:2002nn} for more discussion on their properties).
we also note that the mass dimensions of $\kappa_{10}$, $\Omega$, and $G_3$ are given by $-8$, $-3$, and $-2$, respectively, which is consistent with the mass dimensions of $V_{\rm flux}$ and $W$ given by $4$ and $-5$, respectively.

 Using $2\kappa_{10}^2=g_s^2 \ell_s^8/(2\pi)$ and \eqref{eq:msmpl}, $V_{\rm flux}$ can be rewritten as 
 \dis{V_{\rm flux}=(g_s {\cal V}_0)m_s^6 m_{\rm Pl}^{10}\Big(\frac{m_{\rm Pl}}{m_s}\Big)^6 \sum_{a,b}K^{a\overline{b}} D_aW_{\rm GVW} D_b\overline{W}_{\rm GVW}, }
 from which we find that the superpotential of mass dimension $3$ is given by
 \dis{W=(g_s {\cal V}_0)^{1/2}m_s^3 m_{\rm Pl}^5\Big(\frac{m_{\rm Pl}}{m_s}\Big)^3 \int\Omega\wedge G_3 = \frac{g_s^{3/2}}{\sqrt{4\pi}}\frac{m_{\rm Pl}^6}{\ell_s^2}\Big(\frac{m_{\rm Pl}}{m_s}\Big)^3 \int\Omega\wedge G_3.\label{eq:GVW}}
 The first expression is often convenient since   the coefficient is simply given by $(g_s{\cal V}_0)^{1/2}m_{\rm Pl}^8$  \cite{Bento:2021nbb}.
When we are interested in the EFT below the masses of  the  K\"ahler moduli other than the overall volume modulus, the factor $(m_{\rm Pl}/m_s)^3$ is cancelled by $(m_s/m_{\rm Pl})^3$ in $e^{K/(2m_{\rm Pl}^2)}$, which comes from the last term in \eqref{eq:Kahlergen}.
On the other hand, since the   $G_3$-flux is quantized in units of $\ell_s^2$ (hence has the mass dimension $-2$) and $\Omega$ contains the complex structure moduli which is written in units of   some length scale, say, $m_{\rm Pl}^{-1}$ or $m_s^{-1}$, the second expression is also useful \cite{Conlon:2005ki}.

\renewcommand{\theequation}{\Alph{section}.\arabic{equation}}


\end{document}